\documentclass[acmtog]{acmart}
\acmSubmissionID{1711}

\usepackage{colortbl}
\usepackage{subcaption}
\usepackage{multirow}
\usepackage[linesnumbered,ruled,vlined]{algorithm2e}

\definecolor{tbf}{rgb}{1, 0.7, 0.7} %
\definecolor{tbs}{rgb}{1, 0.85, 0.7} %
\definecolor{tbt}{rgb}{1, 1, 0.7} %
\definecolor{tbn}{rgb}{1, 1, 1} %
\definecolor{gray}{rgb}{0.85, 0.85, 0.85} %
\definecolor{BurntOrange}{HTML}{F7921D}

\SetAlFnt{\small}
\SetAlCapFnt{\small}
\SetAlCapNameFnt{\small}
\SetAlCapHSkip{0pt}

\acmJournal{TOG}
\copyrightyear{2026}
\acmYear{2026}
\setcopyright{cc}
\setcctype{by}
\acmConference[SIGGRAPH Conference Papers '26]{Special Interest Group on Computer Graphics and Interactive Techniques Conference Conference Papers}{July 19--23, 2026}{Los Angeles, CA, USA}
\acmBooktitle{Special Interest Group on Computer Graphics and Interactive Techniques Conference Conference Papers (SIGGRAPH Conference Papers '26), July 19--23, 2026, Los Angeles, CA, USA}
\acmDOI{10.1145/3799902.3811229}
\acmISBN{979-8-4007-2554-8/2026/07}

\begin{document}
\title{Scalable GPU Construction of 3D Voronoi and Power Diagrams}

\author{Bernardo Taveira}
\orcid{0009-0006-4592-2289}
\authornote{Both authors contributed equally to the paper.}
\affiliation{%
 \institution{Zenseact}
 \country{Sweden}
}
\affiliation{
  \institution{Chalmers University of Technology}
  \country{Sweden}
}
\email{bernardo.taveira@chalmers.se}

\author{Carl Lindström}
\orcid{0009-0006-3563-8946}
\authornotemark[1]
\affiliation{%
 \institution{Zenseact}
 \country{Sweden}
}
\affiliation{
  \institution{Chalmers University of Technology}
  \country{Sweden}
}
\email{carl.lindstrom@chalmers.se}

\author{Maryam Fatemi}
\orcid{0009-0003-4513-6737}
\affiliation{%
 \institution{Zenseact}
 \country{Sweden}
}
\email{maryam.fatemi@zenseact.com}

\author{Lars Hammarstrand}
\orcid{0000-0001-5676-1392}
\affiliation{
  \institution{Chalmers University of Technology}
  \country{Sweden}
}
\email{lars.hammarstrand@chalmers.se}

\author{Fredrik Kahl}
\orcid{0000-0001-9835-3020}
\affiliation{
  \institution{Chalmers University of Technology}
  \country{Sweden}
}
\email{fredrik.kahl@chalmers.se}

\renewcommand{\shortauthors}{Taveira, Lindström, et al.}

\begin{teaserfigure}
    \centering
    \includegraphics[width=\textwidth]{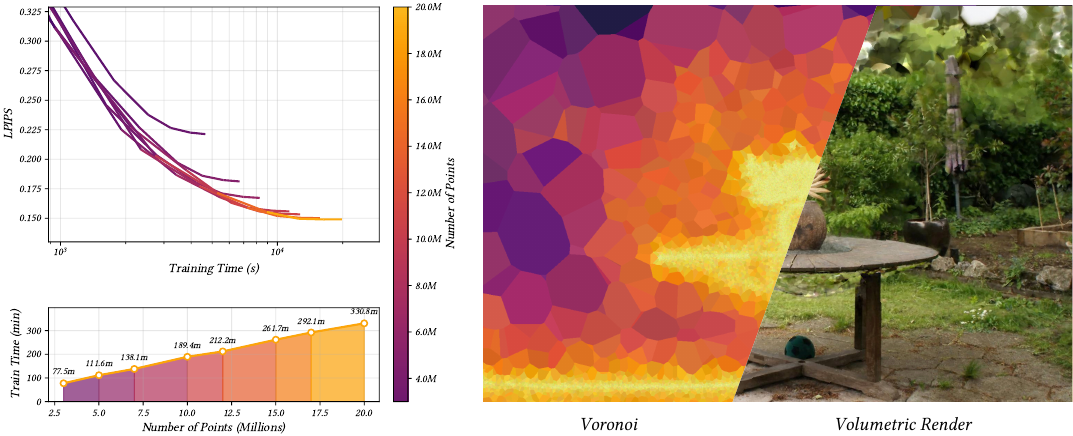}
    \caption{\textbf{Scalability and perceptual quality. } Our efficient construction of 3D Voronoi diagrams enables mesh-based neural rendering to scale to 20 million cells, and reveals a strong correlation between model capacity and perceptual quality. We apply our method to the Radiant Foam~\cite{Govindarajan_2025_ICCV} on the challenging ``Garden'' scene from the mip-NeRF 360 dataset~\cite{barron2022mip}. \textbf{(Left)} The bottom plot shows that total training time (wall-clock, including all optimization steps) increases approximately linearly with the point-count limit. The top plot tracks the evolution of the LPIPS perceptual error across multiple training runs with different point budgets, demonstrating how higher point capacities allow for better visual convergence during training. \textbf{(Right)} A split view of the rendered image and a 2D slice of the Voronoi mesh, color-coded to reflect cell size, highlights the massive and highly varying cell density required to model this scene.}
    \Description{A composite figure with two line plots on the left and a split-view image on the right. The top-left plot shows LPIPS perceptual error decreasing over training time for several runs with different point budgets, with larger point budgets reaching lower LPIPS values. The bottom-left plot shows total training wall-clock time increasing approximately linearly with the point-count limit, from a few million up to 20 million points. The right side shows a photorealistic rendering of an outdoor garden scene next to a 2D slice of its Voronoi mesh, where polygonal cells are color-coded by size, revealing dense fine cells in detailed regions and larger cells in smoother areas.}
    \label{fig:cover}
\end{teaserfigure}
\begin{abstract}

Voronoi diagrams, and their more general weighted counterpart, power diagrams, are fundamental geometric constructs with wide-ranging applications in biology, physics simulation, and computer graphics. Recently, they have gained renewed attention in mesh-based neural rendering. Despite being extensively studied, the construction of 3D Voronoi diagrams for large-scale point sets remains computationally expensive, limiting their adoption in large-scale applications. Existing CPU-based approaches typically rely on computing its dual, the Delaunay tetrahedralization, but are prohibitively slow for large diagrams, while GPU-based methods either struggle to scale efficiently to large point sets or assume homogeneous point distributions. The weighted case, power diagrams, is even less explored in this context. Existing approaches are typically tailored to the application at hand, assuming homogeneous point distributions and small weight variations, making them unsuitable for general use in more complex heterogeneous data.

In this paper, we present a highly parallelizable GPU algorithm for the fast construction of large-scale 3D Voronoi and power diagrams. Our approach constructs each convex cell from a weighted 3D point by progressively clipping an initial cell volume against bisecting planes induced by candidate neighboring points. To efficiently identify candidate neighbors under arbitrary spatial distributions, we introduce a culling criterion based on directional geometric bounds of the evolving cell, combined with a hierarchical best-first traversal of bounding volumes.

We achieve performance on par with state-of-the-art Delaunay tetrahedralization methods on small and moderate problem sizes, while exhibiting robust scalability to large point sets and diverse spatial distributions. Moreover, our method naturally generalizes to power diagrams without additional assumptions. To facilitate reproducibility and future research, we release our source code, see {\color{BurntOrange}\url{https://research.zenseact.com/publications/paragram}}.
\end{abstract}




\maketitle

\section{Introduction}

Voronoi diagrams and their dual, Delaunay triangulations, stand as pillars of computational geometry. For decades, these have been a fundamental concept to computer graphics, enabling mesh generation, fluid simulation, collision detection and surface reconstruction~\cite{aurenhammer1991voronoi, brochu2010matching, shewchuk2002delaunay}. The generalization to weighted points---known as power diagrams and weighted Delaunay triangulation---further extends their utility, enabling the modeling of poly-disperse aggregates and volume preserving partitions. Extensive research, particularly for unweighted cases, has produced reliable and precise methods for computing these geometric structures.

The advent of differentiable rendering has shifted the computational demands from the rendering task into the optimization of the underlying representation. Recent mesh-based neural rendering methods like Radiant Foam \cite{Govindarajan_2025_ICCV} and Radiance Meshes \cite{mai2025radiancemeshesvolumetricreconstruction} demonstrate high quality results and valuable applications by representing 3D space as a mutable volumetric mesh, derived from Voronoi or Delaunay topology. However, these methods introduce a constraint that geometric solvers were never designed to meet: computing diagrams of massive scale (exceeding $10^6$ sites) within iterations of an optimization problem.
As we show in our experiments, the mesh generation frameworks currently employed face scalability and generality limitations that constrain their broader adoption.

This computational bottleneck now limits the scaling of volumetric neural rendering methods. 
While CPU-based solvers are well-studied and precise (e.g.,~\cite{cgal:eb-25b}), they lack the efficiency to scale. GPU alternatives remain underexplored, with existing methods exhibiting stability issues, failing to scale beyond a few million points, or relying on restrictive assumptions on the point distribution.
As the compute hardware improves and neural rendering pushes towards higher resolutions, finer details and larger scenes, the inability to efficiently generate Delaunay and Voronoi meshes becomes the limiting factor on visual quality.

We address this challenge by introducing a highly parallelizable, unified framework for computing 3D power diagrams, weighted Delaunay, and their special cases, Voronoi and Delaunay. Our method is designed to tackle large-scale problems beyond the degree used by current applications. By decoupling cell computation into independent threads, designing a directional culling criteria and utilizing a bounding volume hierarchy tree (BVH) for rapid geometric search and traversal, our algorithm harnesses the massive parallel capabilities of current and future GPUs.

Despite being a general-purpose algorithm, we demonstrate its potential in the context of neural rendering. By replacing the Radiant Foam Voronoi generation method with our drop-in replacement, we effectively remove the computational bottleneck. This allows us to scale the explicit representation from the previous practical limit by 5x, to over 20 million points. We show that this increased capacity alone directly results in improved reconstruction quality.

\section{Related work}

We review prior work on the construction of Voronoi and power diagrams, organized by algorithmic strategy, and conclude with a discussion of mesh-based neural rendering, the application driving our scalability requirements.

\subsection{Voronoi and power diagram construction}

The construction of Voronoi diagrams, power diagrams, and their Delaunay duals is a long-studied problem in computational geometry. Existing solvers fall into three broad algorithmic approaches: incremental insertion, convex-hull lifting, and cell-oriented clipping.

\paragraph{Incremental insertion.} 
Incremental algorithms build the diagram one site at a time, locally retriangulating the cavity of conflicting simplices. The classical Bowyer--Watson method~\cite{Bower81, Watson81} underpins mature CPU libraries such as CGAL~\cite{cgal:eb-25b} and Geogram~\cite{geogramgithub}, both of which ship multi-core implementations. Parallel formulations of this approach have also been the subject of dedicated research~\cite{ParDel03}. A carefully engineered multi-threaded implementation by Marot et al.~\shortcite{marot2019one} tetrahedralizes three billion points on a single workstation in under a minute, and their HXT library~\cite{marot2020quality} represents the current CPU state of the art. On the GPU, gDel3D~\cite{cao2014gpu} adapts this strategy to the massive parallelism the hardware permits, but its resource footprint limits scalability to a few million sites on consumer hardware.

\paragraph{Convex-hull lifting.}
Lifting sites onto a paraboloid in $\mathbb{R}^{d+1}$ reduces Voronoi and power diagram construction to a lower-convex-hull problem~\cite{aurenhammer1987power}. This formulation naturally accommodates the weighted case, and is used by the Quickhull implementation QHull~\cite{QHull96} and the Delaunay routines of SciPy~\cite{2020SciPy-NMeth}, which rely on QHull internally. However, the added dimension makes it uncompetitive for the large 3D point sets we target.

\paragraph{Cell-oriented clipping.}
Cell-oriented methods construct each cell independently by progressively clipping an initial convex volume against bisecting half-spaces induced by neighboring sites. This formulation was popularized by the CPU library Voro++~\cite{rycroft2009voro++}, later extended to multi-threaded execution~\cite{voro2023extension}. A key concept for efficient cell clipping is the radius of security criterion of L\'evy and Bonneel~\shortcite{levy2013variational}, which guarantees that the cell has been clipped by enough neighbors, allowing early termination of per-cell construction. Sainlot et al.~\shortcite{sainlot2017restricting} build on this idea, proposing corner validation as an alternative termination criterion. Ray et al.~\shortcite{MeshlessVor18} and the subsequent work of Basselin et al.~\shortcite{10.1111:cgf.142610} brought the cell-oriented strategy to the GPU, targeting meshless volume integration. Their approach assumes a near-uniform distribution of sites. Neighbor search is driven by a $k$-nearest-neighbor heuristic and terminates at a fixed multiple of the current cell radius. These assumptions break down in the heterogeneous and weighted settings we target, where local density, and thus the required search radius, varies by orders of magnitude across the domain.

Our method builds on Basselin et al.~\shortcite{10.1111:cgf.142610}. We preserve the per-cell independence that makes this formulation well suited to massive parallelism, and replace the isotropic radius criterion with a \emph{directional} culling criterion derived from the evolving axis-aligned bounds of each cell, combined with a hierarchical best-first traversal over a BVH augmented for power-distance queries. Together, these components enable a GPU-native construction that scales to tens of millions of sites under arbitrary spatial and weight distributions.

\subsection{Mesh-based view synthesis}
Research on novel view synthesis was significantly accelerated by Neural Radiance Fields (NeRF)~\cite{mildenhall2021nerf}, which represents scenes as continuous volumetric radiance fields optimized via differentiable volume rendering. Subsequent work has focused on improving rendering efficiency by adopting explicit scene representations and point-based rasterization, most notably through 3D Gaussian Splatting (3DGS)~\cite{kerbl20233Dgaussians}.

Given the foundational role of meshes in computer graphics, several approaches have explored the use of meshes for both NeRF-style implicit fields~\cite{wang2021neus, yariv2021volume, yariv2023bakedsdf} and 3DGS-style explicit representations~\cite{choi2024meshgs, gao2024real,lin2024direct}. More recent work has investigated differentiable, mesh-based scene representations that explicitly discretize space. Radiant Foam \cite{Govindarajan_2025_ICCV} represents scenes as an optimizable 3D Voronoi diagram. To support accurate ray-tracing, it requires frequent updates of cell adjacencies during training via Delaunay triangulation. Similarly, Radiance Meshes~\cite{mai2025radiancemeshesvolumetricreconstruction} partitions space into tetrahedral cells derived from gDel3D's~\cite{cao2014gpu} Delaunay tetrahedralization.

These methods typically initialize cell sites using structure-from-motion~\cite{schoenberger2016sfm}, then optimize their positions via gradient descent on photometric reconstruction losses, a process requiring up to 20k optimization steps. Heuristics are also employed to densify and prune the representation throughout training.

While these approaches enable explicit control over scene discretization, they critically depend on efficient triangulation algorithms, as connectivity must be updated frequently during optimization. Moreover, visual fidelity scales with the number of sites, motivating the need for larger geometric meshes and scalable construction methods.

\section{Preliminaries}

To understand our method, we begin by formally defining the geometric structures that underlie it.
Let $P = \{p_1, \ldots, p_n\} \subset \mathbb{R}^d$ be a set of $n$ points, each with an associated scalar weight $w_i \in \mathbb{R}$.
The \emph{power diagram} of $P$ partitions space into convex cells according to the power distance, where each cell site $p_i$ is

\begin{equation} \label{eq:power_diagram_cell}
    C_i = \{ x \in \mathbb{R}^d \mid \|x - p_i\|^2 - w_i \le \|x - p_j\|^2 - w_j,\ \forall j \ne i \}.
\end{equation}

When all weights are equal, the problem is reduced to the standard Voronoi diagram, which partitions space by nearest neighbors. The sites $p_i$ and $p_j$ are considered adjacent if their cells share a polygonal face. The resulting adjacency graph defines the connectivity of the dual \emph{regular} (weighted) Delaunay triangulation.

Equation~\ref{eq:power_diagram_cell} reveals a key property for parallel construction. Each power cell $C_i$ is computationally independent, defined solely by the intersection of half-spaces derived from its neighbors. This independence implies that the construction of the power diagram of $P$ can be decomposed into $n$ local clipping tasks. However, computational efficiency depends on identifying relevant neighbors from the large set $P$ that contribute to the cell boundary.

\begin{figure*}[t]
  \centering
  \begin{subfigure}[t]{0.32\textwidth}
    \centering
    \includegraphics[width=\linewidth]{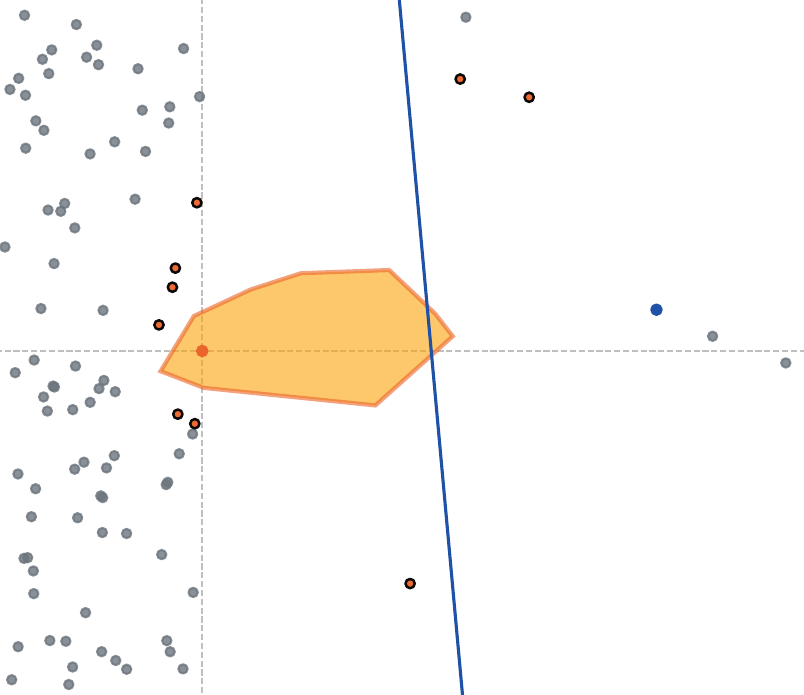}
    \caption{Convex cell clipping by a bisecting plane. The blue site induces a plane that intersects the current cell (orange) and therefore must be considered during clipping.}
    \Description{Schematic of a central seed with its orange convex cell and a blue candidate site whose bisecting plane (dashed line) crosses the cell, indicating it must be considered for clipping.}
    \label{fig:geo_bounds:a}
  \end{subfigure}
  \hfill
  \begin{subfigure}[t]{0.32\textwidth}
    \centering
    \includegraphics[width=\linewidth]{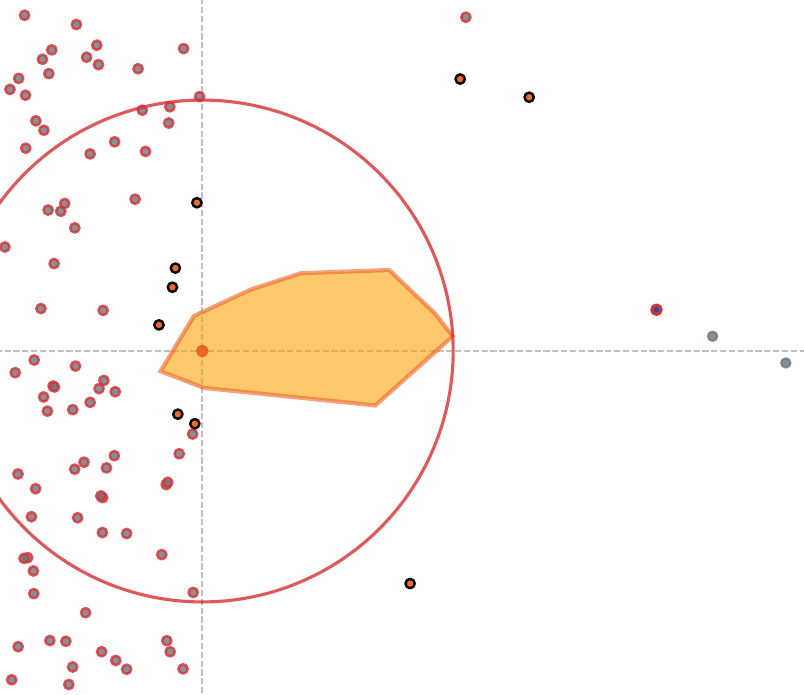}
    \caption{Isotropic culling using the maximum cell radius. All sites within the global radius are retained as candidates (red), including many whose bisecting planes cannot intersect the cell.}
    \Description{The same orange cell overlaid with a circle of radius equal to the maximum cell extent; all neighbors inside the circle are marked red as retained candidates, including many that cannot actually clip the cell.}
    \label{fig:geo_bounds:b}
  \end{subfigure}
  \hfill
  \begin{subfigure}[t]{0.32\textwidth}
    \centering
    \includegraphics[width=\linewidth]{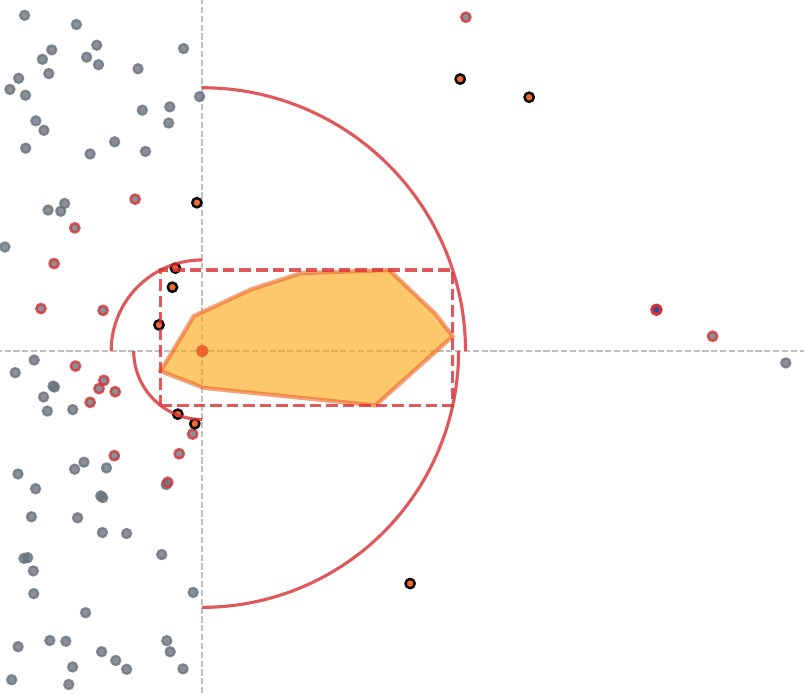}
    \caption{Directional culling using AABB-based geometric bounds yields a significantly tighter candidate set (red).}
    \Description{The same orange cell with its axis-aligned bounding box and direction-dependent radii forming a tighter envelope, retaining far fewer red candidate neighbors than the isotropic bound.}
    \label{fig:geo_bounds:c}
  \end{subfigure}
  \caption{Effect of geometric bounds on neighbor culling during convex cell construction. The current cell is shown in orange, with previously accepted neighbor sites highlighted. (a) A site whose bisecting plane intersects the cell (blue) must be selected in the candidate set for cell clipping. (b) An isotropic bound based on the maximum cell radius correctly includes the site, but also admits many irrelevant candidates. (c) Directional geometric bounds derived from an AABB of the cell provide a tighter criterion, reducing the number of unnecessary neighbors while preserving correctness.}
  \Description{Three-panel schematic comparing neighbor-culling bounds: (a) a relevant blue candidate whose bisecting plane crosses the orange cell, (b) an isotropic circular bound that retains many unnecessary red candidates, and (c) a tighter direction-dependent AABB-based bound that retains far fewer.}
  \label{fig:culling_criteria}
\end{figure*}
\section{Method}
Given the set of weighted points $P$, our goal is to compute the power diagram where each point's cell is given by Eq.~\ref{eq:power_diagram_cell} and forms a convex polyhedron defined by the intersection of half-spaces induced by bisecting planes with neighboring cells. Note that the special case of equal weights reduces to the Voronoi diagram and its dual, the Delaunay triangulation.

To this end, we present a new GPU-accelerated algorithm for computing 3D power diagrams at scale. Our method combines three key components:
\begin{itemize}
    \item [(1)] a convex cell clipping algorithm to construct individual diagram cells (Sec.~\ref{sec:convex_cell_clipping}),
    \item [(2)] a directional culling criterion that efficiently identifies relevant neighbor sites (Sec.~\ref{sec:culling}), and
    \item[(3)] a hierarchical spatial data structure that enables rapid neighbor queries (Sec.~\ref{sec:hierarchical}).
\end{itemize}
Our key technical innovation efficiently discards large volumes of irrelevant neighbor sites using tight geometric bounds and a hierarchical data structure, while guaranteeing correctness.

\subsection{Convex cell clipping}\label{sec:convex_cell_clipping}
The core of our algorithm lies in computing each cell independently as the intersection of a set of half-spaces, following \cite{10.1111:cgf.142610}. In a power diagram, the half-spaces of a cell are defined by the bisection planes between a site and its neighbors. Consequently, given the adjacency of a site, the corresponding convex polyhedron can be constructed directly via half-plane intersection. Naively, if a cell is iteratively clipped by all bisecting planes induced by other sites in the point set, and the sites whose planes contribute to the final boundary are recorded, the resulting convex polyhedron and its adjacency are recovered.

The clipping procedure follows \cite{10.1111:cgf.142610}. For each candidate neighbor, the vertices lying outside the corresponding half-space are removed. The intersection between the bisecting plane and the current convex cell is then computed, producing a polygonal boundary whose vertices are added to the cell. If the bisecting plane lies entirely outside the current cell, the intersection is empty and the cell remains unchanged. See the supplementary material for a detailed description. Fig.\ref{fig:culling_criteria}(a) demonstrates the complementary case where a non-empty intersection triggers clipping. To achieve optimal performance, clipping should begin with neighbors that are most likely to contribute to the final cell. Early reduction of the cell volume increases the likelihood that subsequent clipping operations result in no modification and can be skipped.

\subsection{Geometric bounds and culling criteria}\label{sec:culling}
The brute-force cell sculpting procedure described in Sec.~\ref{sec:convex_cell_clipping} guarantees the correct diagram construction, but becomes prohibitively expensive for large point sets if all points are treated as potential neighbors. As the cell is progressively clipped, however, many neighbors can be rejected early if their bisecting planes are too distant to intersect the current cell. Since exact intersection tests are costly, we instead rely on conservative geometric bounds.

For a site $p_i \in \mathbb{R}^3$ with weight $w_i \in \mathbb{R}$ and a candidate neighbor $p_j \in \mathbb{R}^3$ with weight $w_j \in \mathbb{R}$, the (power) distance from $p_i$ to their bisecting plane is
\begin{equation}
   d_{ij} = \frac{|| p_i - p_j ||^2 + w_i - w_j }{2 ||p_i - p_j ||}.
\end{equation}
Given a geometric bound $r_i$ on the extent of the cell $C_i$, the candidate neighbor $C_j$ can be safely discarded if $d_{ij} > r_i$. Provided that $r_i$ upper-bounds the distance to all vertices relevant to that neighbor, this criterion preserves correctness. 

\subsubsection{Directional culling} 
Prior work \cite{10.1111:cgf.142610} employs an isotropic bound defined as the maximum distance from the cell site to any current cell vertex, see Fig.\ref{fig:culling_criteria}(b). While effective for roughly uniform point distributions, this bound becomes overly conservative for more complex configurations, when the cell's spatial extent varies significantly across directions.

We address this limitation with a \emph{directional geometric bound} that depends on the relative position of the candidate neighbor. The key observation is that a bisecting plane can only intersect a subset of the cell faces depending on the direction vector $p_j-p_i$. By restricting the bound based on those faces, we obtain a tighter and more informative culling criterion. 

Specifically, we define a \emph{directional cell radius} using the distances from the cell site to the corners of an axis-aligned bounding box (AABB) enclosing the current cell vertices, Fig.\ref{fig:culling_criteria}(c). The appropriate bound for a candidate neighbor is determined by identifying the octant of space in which the neighbor $p_j$ lies relative to the cell site $p_i$. This octant uniquely determines which subset of AABB corners can possibly intersect with the corresponding bisecting plane.

The directional radius is the maximum distance to these corners, giving a conservative upper bound on the cell's extent toward the candidate neighbor. This significantly reduces unnecessary clipping while retaining the efficiency of a max over a small, fixed set.

\subsubsection{Bounding-volume culling} \label{sec:volume_culling}
Evaluating the directional culling criterion for every individual site can still be inefficient for large point sets. Prior work \cite{10.1111:cgf.142610} mitigates this by iteratively querying the $k$ nearest neighbors in a growing search radius, which works well for homogeneous point distributions and small weight differences between cells. In contrast, we make no assumptions on point distributions, and instead formulate a general strategy for discarding entire volumes of candidates.

We extend directional geometric bounds to spatial groups of candidates by testing entire axis-aligned bounding volumes at once. Specifically, we determine whether any point $p_j$ within a volume $B= [ \, b_{min},b_{max} \, ]$, where $b_{min},b_{max} \in \mathbb{R}^3$, could clip the current cell defined by $p_i$. Since we assume no symmetry, size, or placement of the volume, the test naturally supports directional discarding and applies to arbitrary spatial partitions.

Each volume $B$ is associated with the maximum weight $w_{max} = \max_j w_j$ of any site it contains. This represents a worst case, as higher weights induce bisecting planes closer to the cell site. We conservatively assume the closest potential candidate lies on the surface of $B$ at the location minimizing its distance to $p_i$, giving a lower bound on all bisecting-plane distances $d_{ij}$ for points $p_j$ in $B$.

Because $B$ may span multiple octants relative to $p_i$, we compute the directional radius for each occupied octant and take their maximum as a conservative bound. This upper-bounds the cell extent in all directions from which a candidate in $B$ could affect it.

Finally, we compare the bisecting-plane distance induced by the closest possible site on the boundary of $B$, using the maximum weight $w_{max}$, against the upper-bound $r_i$. If $d$ is the distance between the closest possible boundary point of $B$ and $p_i$, then
$$
   d_{ij} = \frac{|| p_i - p_j ||}{2} + \frac{w_i - w_j }{2 ||p_i - p_j ||} \ge d/2 + \frac{w_i - w_{max} }{2 d}
$$
provided $w_i \le w_{max}$, and otherwise,  $d_{ij} \ge d/2$ is a valid lower bound.
If this lower bound on $d_{ij}$ exceeds $r_i$, no candidate in $B$ can clip the current cell, and the entire volume is discarded.

\subsection{Hierarchical neighbor search}\label{sec:hierarchical}
\begin{algorithm}[t]
\caption{Best-First Cell Neighbor Traversal}
\label{alg:best-first}
\SetKwInOut{Input}{Input}
\SetKwFunction{IsLeaf}{IsLeaf}
\DontPrintSemicolon
\Input{BVH tree $\mathcal{T}$, Seed point $P$, Initial Radii $R$}
$stack \leftarrow \emptyset$\;
$node \leftarrow \mathcal{T}.root$\;
\While{\textbf{true}}{
    \tcp{Traverse down until a leaf or dead-end is reached}
    \While{\textbf{not} \IsLeaf{$node$}}{
        $n_0, n_1 \leftarrow node.children$\;
        $r_0 \leftarrow \textsc{ComputeDirectionalRadius}(R, P, n_0.bounds)^2$\;
        $r_1 \leftarrow \textsc{ComputeDirectionalRadius}(R, P, n_1.bounds)^2$\;
        $d_0 \leftarrow \textsc{NodeSqrDist}(n_0)$; \quad $d_1 \leftarrow \textsc{NodeSqrDist}(n_1)$\;
        
        $\delta_0 \leftarrow d_0 - r_0$; \quad $\delta_1 \leftarrow d_1 - r_1$\;

        \If{$\min(\delta_0, \delta_1) > 0$}{
            $node \leftarrow \text{NULL}$ \tcp*{Both children too far}
            \textbf{break}\;
        }

        \tcp{Traverse near child, push far child if potentially valid}
        \eIf{$\delta_0 < \delta_1$}{
            $node \leftarrow n_0$; \quad $far \leftarrow \{n_1, d_1\}$; \quad $\delta_{far} \leftarrow \delta_1$\;
        }{
            $node \leftarrow n_1$; \quad $far \leftarrow \{n_0, d_0\}$; \quad $\delta_{far} \leftarrow \delta_0$\;
        }

        \If{$\delta_{far} \leq 0$}{
            $stack.\text{Push}(far)$\;
        }
    }
    \If{$node \neq \text{NULL}$}{
        \tcp{We reached a valid leaf}
        $R \leftarrow \textsc{ProcessLeaf}(node.prims, R)$\;
    }
    \tcp{Backtrack: Pop next candidate}
    \While{\textbf{true}}{
        \If{$stack$ is empty}{
            \Return\;
        }
        $\{idx, dist\} \leftarrow stack.\text{Pop}()$\;
        
        $r_{new} \leftarrow \textsc{ComputeDirectionalRadius}(R, P, \mathcal{T}[idx].bounds)^2$\;
        
        \tcp{Check if node is still valid with potentially shrunk $R$}
        \If{$dist - r_{new} \leq 0$}{
            $node \leftarrow \mathcal{T}[idx]$\;
            \textbf{break} \tcp*{Resume traversal}
        }
    }
}
\end{algorithm}

Efficiently identifying the subset of sites that define the cell boundary requires a spatial acceleration structure that supports rapid culling of distant candidates. Furthermore, the amount of clips required to reach the final cell polyhedra is highly dependent on the order in which planes are processed. Processing the nearest neighbors first can rapidly shrink the cell, tightening the bounding box and allowing us to discard large portions of the search space early.

To partition the sites, we employ a bounding volume hierarchy (BVH). This structure inherently supports hierarchical, multi-scale culling: a single test at a high-level node can discard vast regions containing millions of sites, while deeper traversals allow precise culling of smaller subsets. This spatial adaptivity is crucial for distributions of drastically varying densities, where uniform grids or voxelization approaches \cite{10.1111:cgf.142610} struggle to balance skipping large empty volumes with processing high-density clusters. To support power diagrams, we augment each BVH node to not only store its volumetric bounds but also the maximum weight of any site in its subtree, enabling the culling test in Sec.~\ref{sec:volume_culling}.

Standard BVH traversals usually employ a depth-first search (DFS) using a LIFO (last-in-first-out) stack. However, DFS does not ensure that nodes are visited in order of proximity, leading to wasted cell clipping operations against distant sites that are later occluded and discarded as adjacent. Instead, we employ a best-first search, using a local priority queue ordered by distance. This approach prioritizes proximity, ensuring that we clip the most restrictive planes early in the process, rapidly shrinking the cell.

During the traversal and clipping process we maintain and update an axis-aligned bounding box (AABB) of the current partially clipped cell. As described in Sec.~\ref{sec:culling}, this bounding box is used to determine if any point within a volume could potentially clip the cell bounding box. The same signed distance function is employed to pick the closest of the node's children. By using this signed function instead of the euclidean distance to the site seed, we prioritize the neighbors most likely to induce a cut on the current bounding volume. The signed distance metric effectively measures how deep a candidate site can potentially penetrates the current validity region of the cell. By prioritizing based on this penetration depth, we ensure that the algorithm processes the nodes most likely to clip the cell in a significant manner. This strategy maximizes the rate of volume reduction, as the most intrusive planes are clipped immediately. Pseudo-code of this best-first neighbor traversal is provided in Alg.~\ref{alg:best-first}.

\section{Experiments}
We evaluate our method across synthetic datasets and a real-world neural rendering application to demonstrate both computational efficiency and practical applicability. Our experiments are designed to evaluate our method's performance across different spatial distributions, how it scales with an increasing number of points, and its effectiveness in existing applications.
We describe the datasets used for our experiments in Sec.~\ref{sec:experiments:datasets}, and present the corresponding results in Sec.~\ref{sec:experiments:results}. Finally, we demonstrate how our method can scale up neural rendering applications in Sec.~\ref{sec:experiments:scaling_neural_rendering}.

\subsection{Datasets} \label{sec:experiments:datasets}
\paragraph{Synthetic test cases.}
To evaluate the computational efficiency of our method across different point set distributions, we generate three synthetic test configurations. All configurations sample points within the domain $\Omega = [-10, 10]^3$. The first configuration samples points uniformly:
\begin{equation}
    p_i \sim \mathcal{U}(\Omega).
\end{equation}
This represents generic spatial distributions with no inherent structure but varying density.

The second configuration samples points from multiple Gaussian clusters to simulate scenarios with local density variations and large empty regions. We first sample $K$ cluster centers uniformly in $\Omega$, then distribute points evenly among clusters:
\begin{equation}
    c_j \sim \mathcal{U}(\Omega), \quad p_i \sim \mathcal{N}(c_j, \sigma^2\mathbf{I}),
\end{equation}
where $\sigma=0.1$ and points are clamped to $\Omega$. We evaluate with $K \in \{5, 10\}$ clusters.

In the third test configuration, we sample points with a linear density gradient along the x-axis using inverse transform sampling:
\begin{align}
    p(x) \propto (x - a), \quad x \in [a, b] \\
    \Rightarrow x = a + (b-a)\sqrt{u}, \quad u \sim \mathcal{U}(0,1) \\
    y,z \sim \mathcal{U}(a, b)
\end{align}
where $[a,b]=[-10, 10]$. This configuration produces higher point density at $x=10$ and lower density at $x=-10$.

For each synthetic test configuration, we evaluate performance with point counts ranging from $0.1$M to $15$M, spanning from moderate to large-scale problems. Fig.~\ref{fig:synthetic_cases_collage} shows a visualization of the synthetic test cases.

\begin{figure*}
    \centering
    \includegraphics[width=1.0\linewidth]{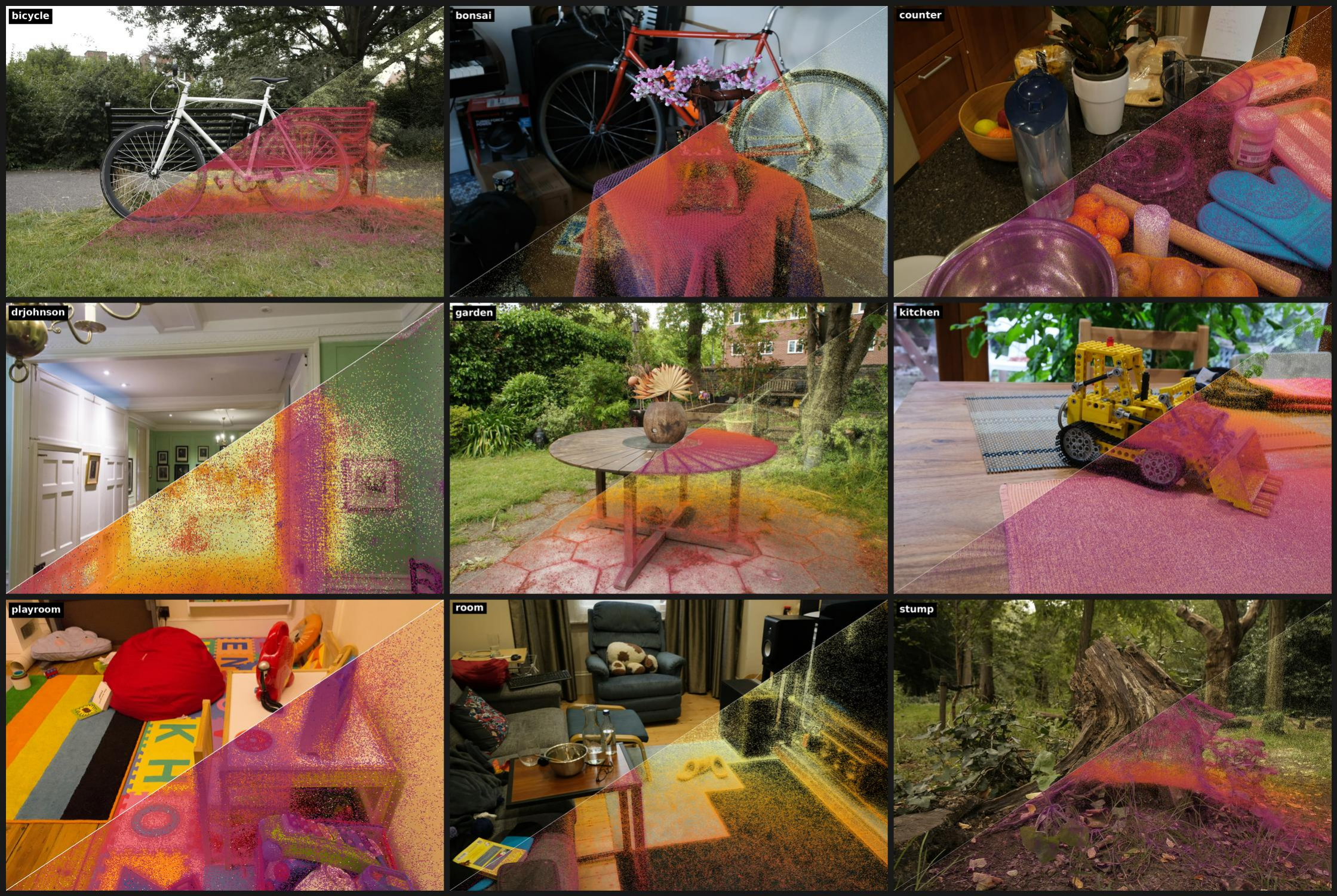}
    \caption{
    Images from the real-world test cases (described in Sec.~\ref{sec:experiments:datasets}), with point sets projected to illustrate the density and spatial spread of the data. The scenes span a range of indoor and outdoor environments with complex spatial distributions and point counts ranging from 2M to 4.2M.
    }
    \Description{A collage of photographs from real-world scenes used as benchmarks (the seven Mip-NeRF 360 scenes such as bicycle, garden, stump, kitchen, bonsai, room, counter, plus two Deep Blending scenes drjohnson and playroom). Each image is overlaid with the projected Voronoi seed points from a trained Radiant Foam checkpoint, visualizing how points concentrate densely on visible surfaces and foreground objects while being more sparsely distributed in background regions, illustrating the highly non-uniform spatial distributions of these point sets.}
    \label{fig:dataset_collage}
\end{figure*}

\paragraph{Real-world test cases.}
To evaluate our method on realistic point distributions that arise in practical applications, we use trained checkpoints from Radiant Foam \cite{Govindarajan_2025_ICCV}, a neural rendering model that uses 3D Voronoi diagrams as its differentiable scene representation. We obtain publicly available checkpoints from the original paper, trained on seven scenes from the Mip-NeRF 360 dataset~\cite{barron2022mip} and two scenes from the Deep Blending dataset~\cite{hedman2018deep}. Each checkpoint contains the seed point positions of the final Voronoi diagram configuration after optimization convergence. These checkpoints provide representative examples of moderately large point sets (ranging from 2M to 4.2M sites, depending on the scene) with complex spatial distributions that emerge from volumetric rendering optimization. These test cases exhibit realistic patterns of local clustering, empty space, and varying density that reflect the underlying scene geometry. Fig.~\ref{fig:dataset_collage} shows a visualization of the real-world test cases.

\paragraph{Power diagrams.}
To validate that our method remains efficient for the more general case of weighted Voronoi diagrams (i.e., power diagrams), we evaluate on both the synthetic and real-world test cases with weighted point sets. We sample weights from a Gaussian distribution:
\begin{equation}
    w_i \sim \mathcal{N} ( 0,  ( \frac{d_\text{nn}^2}{3} )^2 )
\end{equation}
where $d_\text{nn}$ is the median nearest-neighbor distance in the point set. We use identical point positions as in the unweighted experiments to isolate the impact of weights on performance.

\subsection{Results}\label{sec:experiments:results}
We evaluate our algorithm on the datasets presented in Sec.~\ref{sec:experiments:datasets}, measuring computational efficiency on both consumer-grade GPUs (NVIDIA RTX 5090) and enterprise GPUs (NVIDIA H200). We compare against two GPU-based algorithms for 3D Delaunay triangulation, gDel3D~\cite{cao2014gpu} and the implementation from Radiant Foam~\cite{Govindarajan_2025_ICCV}. Further, we include our own CUDA implementation of \cite{10.1111:cgf.142610}, adapted to facilitate explicit mesh construction. Additionally, we evaluate CPU-based implementations from SciPy~\cite{2020SciPy-NMeth}, CGAL~\cite{cgal:eb-25b}, Geogram~\cite{geogramgithub}, Voro++~\cite{rycroft2009voro++} and HXT GmSH~\cite{marot2020quality}. For CPU methods, we use 16 CPU cores from an AMD Epyc 9534. We run three warm-up iterations followed by ten timed runs from which we compute the average runtime. We impose a 300-second timeout limit. All reported timing results are end-to-end; no preprocessing or preallocation is excluded.

\begin{figure*}
    \centering
    \includegraphics[width=0.75\linewidth]{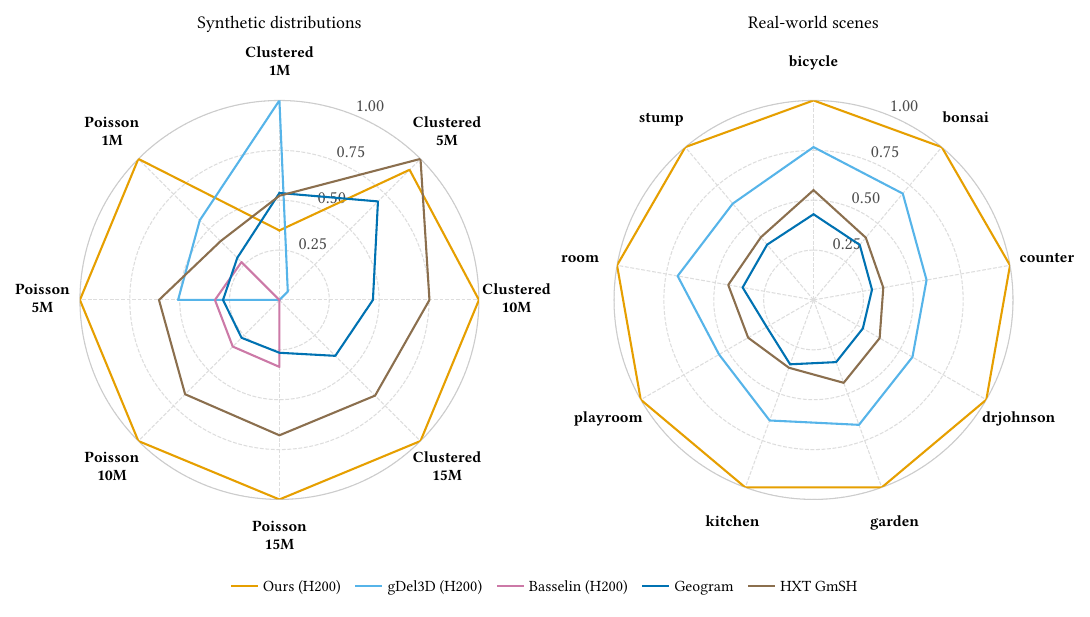}
    \caption{
    Runtime comparison on synthetic and real-world point sets. The outer ring corresponds to the fastest method on that dataset (score = 1). A score of zero indicates that the method failed to complete a valid result. \textit{Poisson} represents the average of the Poisson-distributed test cases (white noise and density gradient). \textit{Clustered} represents the average of the clustered test cases. Our method consistently achieves scores near the outer ring across all distributions, while competitors degrade at higher point counts and non-uniform distributions.
    }
    \Description{Radar (spider) plots with axes for competing methods, normalized so the fastest method on each dataset reaches the outer ring (score 1) and failures are pinned at the center, grouped by dataset category and point-count regime. Our method's polygon stays close to the outer ring across all distributions while baselines collapse toward the center at larger point counts and non-uniform distributions.}
    \label{fig:radio_plot_h200}
\end{figure*}

\paragraph{Computational efficiency.}
For the synthetic test cases, we evaluate point counts of $0.1$M, $0.5$M, $1$M, $2$M, $5$M, $10$M, and $15$M, presenting results in Fig.~\ref{fig:synthetic_cases_collage}. Detailed numerical results for Fig.~\ref{fig:synthetic_cases_collage} are provided in the supplementary material. We also measure runtime on the trained checkpoints from Radiant Foam (described in Sec.~\ref{sec:experiments:datasets}) and present results in Tab.~\ref{tab:timing_results_mipnerf360_joint}. Finally, we show aggregated results of the best performing methods in Fig.~\ref{fig:radio_plot_h200}.

As illustrated in Fig.~\ref{fig:radio_plot_h200}, our algorithm is competitive across all point distributions and scales. Notably, it is the fastest across all point sizes for both the Poisson-distributed cases (density gradient and white noise) and the real-world scenes, and the fastest overall for large point sets ($\geq$5M). 
Among the GPU-baselines, gDel3D achieves the best efficiency on the small clustered point sets, but exhausts memory for larger scenes even on enterprise hardware. On real-world scenes it is on average $\sim 55\%$ slower than our method on the H200 and $\sim 137\%$ slower on the 5090. Our implementation of~\cite{10.1111:cgf.142610} scales to large point sets on Poisson-distributed data, but fails to complete any other test cases due to the limitations of KNN-based search on irregular distributions. HXT is the best-performing CPU-baseline, closely followed by Geogram. HXT performs very close to our method on the clustered data and scales well to larger point sets, but is on average $\sim 55\%$ slower than our method (H200) on the Poisson-distributed data and $\sim 129\%$ slower on real-world scenes.
Overall, our method outperforms both CPU and GPU baselines on both real-world scenes and Poisson-distributed data across all point set sizes, while remaining competitive on the clustered data, particularly for large point sets. Further details on precision and timing breakdown of our method are provided in the supplementary material.

\begin{table}[t]
    \small
    \centering
    \setlength{\tabcolsep}{1.0pt}
    \caption{\textbf{Voronoi diagram creation/Delaunay triangulation}. Runtime results (in seconds) on point sets obtained from Radiant Foam checkpoints trained on the Mip-NeRF 360 dataset. Color coding denotes \colorbox{tbf}{fastest}, \colorbox{tbs}{second fastest}, and \colorbox{tbt}{third fastest}.}
    \Description{Voronoi/Delaunay runtimes in seconds across nine real-world scenes (1.9M--4.2M points) for CPU and GPU baselines, with the three fastest per scene color-coded. Our method is fastest on every scene on both the RTX 5090 and H200, while Voro++ and Basselin* fail on these distributions.}
    \setlength{\tabcolsep}{2.5pt}
    \resizebox{1.0\linewidth}{!}{
    \begin{tabular}{cl ccccccccc}
    \toprule
    & & bicycle & bonsai & counter & drjohnson & garden & kitchen & playroom & room & stump \\
    & & 4.2M & 2.0M & 2.0M & 2.8M & 4.1M & 2.0M & 3.2M & 1.9M & 4.2M \\
    \midrule

    \multirow{6}{*}{\rotatebox{90}{CPU}}
    & SciPy           & \cellcolor{tbn}279.710 & \cellcolor{tbn}121.480 & \cellcolor{tbn}127.273 & \cellcolor{tbn}187.908 & \cellcolor{tbn}260.904 & \cellcolor{tbn}133.174 & \cellcolor{tbn}207.022 & \cellcolor{tbn}119.729 & \cellcolor{tbn}268.345\\
    & Voro++           & \cellcolor{tbn}- & \cellcolor{tbn}- & \cellcolor{tbn}- & \cellcolor{tbn}- & \cellcolor{tbn}- & \cellcolor{tbn}- & \cellcolor{tbn}- & \cellcolor{tbn}- & \cellcolor{tbn}-\\
    & CGAL            & \cellcolor{tbn}20.445 & \cellcolor{tbn}9.591 & \cellcolor{tbn}9.366 & \cellcolor{tbn}13.522 & \cellcolor{tbn}19.205 & \cellcolor{tbn}9.781 & \cellcolor{tbn}15.016 & \cellcolor{tbn}8.877 & \cellcolor{tbn}19.559\\
    & CGAL Parallel            & \cellcolor{tbn}26.986 & \cellcolor{tbn}18.900 & \cellcolor{tbn}97.873 & \cellcolor{tbn}21.249 & \cellcolor{tbn}20.066 & \cellcolor{tbn}57.639 & \cellcolor{tbn}49.721 & \cellcolor{tbn}40.528 & \cellcolor{tbn}40.466\\
    & Geogram   & \cellcolor{tbn}1.593 & \cellcolor{tbn}0.826 & \cellcolor{tbn}0.859 & \cellcolor{tbn}1.244 & \cellcolor{tbn}1.732 & \cellcolor{tbn}0.865 & \cellcolor{tbn}1.370 & \cellcolor{tbn}0.812 & \cellcolor{tbn}1.560\\
    & HXT GmSH   & \cellcolor{tbt}1.243 & \cellcolor{tbt}0.730 & \cellcolor{tbt}0.720 & \cellcolor{tbt}0.928 & \cellcolor{tbt}1.299 & \cellcolor{tbt}0.821 & \cellcolor{tbt}0.980 & \cellcolor{tbt}0.675 & \cellcolor{tbt}1.375\\
    \midrule
    \multirow{4}{*}{\rotatebox{90}{5090}} 
    & Basselin*         & \cellcolor{tbn}- & \cellcolor{tbn}- & \cellcolor{tbn}- & \cellcolor{tbn}- & \cellcolor{tbn}- & \cellcolor{tbn}- & \cellcolor{tbn}- & \cellcolor{tbn}- & \cellcolor{tbn}-\\
    & RF Del         & \cellcolor{tbn}124.929 & \cellcolor{tbn}25.299 & \cellcolor{tbn}24.980 & \cellcolor{tbn}24.201 & \cellcolor{tbn}40.413 & \cellcolor{tbn}24.762 & \cellcolor{tbn}93.342 & \cellcolor{tbn}16.889 & \cellcolor{tbn}44.777\\
    & gDel3D         & \cellcolor{tbs}1.204 & \cellcolor{tbs}0.570 & \cellcolor{tbs}0.592 & \cellcolor{tbs}0.826 & \cellcolor{tbs}1.168 & \cellcolor{tbs}0.604 & \cellcolor{tbs}0.931 & \cellcolor{tbs}0.542 & \cellcolor{tbs}1.197\\
    & Ours     & \cellcolor{tbf}0.570 & \cellcolor{tbf}0.245 & \cellcolor{tbf}0.248 & \cellcolor{tbf}0.331 & \cellcolor{tbf}0.496 & \cellcolor{tbf}0.235 & \cellcolor{tbf}0.359 & \cellcolor{tbf}0.240 & \cellcolor{tbf}0.491\\
    \midrule
    \multirow{4}{*}{\rotatebox{90}{H200}} 
    &Basselin*         & \cellcolor{tbn}- & \cellcolor{tbn}- & \cellcolor{tbn}- & \cellcolor{tbn}- & \cellcolor{tbn}- & \cellcolor{tbn}- & \cellcolor{tbn}- & \cellcolor{tbn}- & \cellcolor{tbn}-\\
    & RF Del         & \cellcolor{tbn}250.688 & \cellcolor{tbn}22.501 & \cellcolor{tbn}24.800 & \cellcolor{tbn}25.142 & \cellcolor{tbn}39.551 & \cellcolor{tbn}25.851 & \cellcolor{tbn}122.038 & \cellcolor{tbn}18.020 & \cellcolor{tbn}43.680\\
    & gDel3D         & \cellcolor{tbs}0.893 & \cellcolor{tbs}0.428 & \cellcolor{tbs}0.444 & \cellcolor{tbs}0.620 & \cellcolor{tbs}0.862 & \cellcolor{tbs}0.462 & \cellcolor{tbs}0.678 & \cellcolor{tbs}0.424 & \cellcolor{tbs}0.896\\
    & Ours     & \cellcolor{tbf}0.684 & \cellcolor{tbf}0.298 & \cellcolor{tbf}0.256 & \cellcolor{tbf}0.355 & \cellcolor{tbf}0.574 & \cellcolor{tbf}0.297 & \cellcolor{tbf}0.371 & \cellcolor{tbf}0.293 & \cellcolor{tbf}0.564\\
    \bottomrule
    \end{tabular}
    }
    \label{tab:timing_results_mipnerf360_joint}
\end{table}

\paragraph{Generalization to power diagrams.}
 As our method generalizes to power diagrams, we validate performance on the weighted version of our datasets, comparing against CPU-based algorithms for Regular Delaunay triangulation from SciPy, CGAL and Geogram. We also include results from our implementation of ~\cite{10.1111:cgf.142610} on the white noise and density gradient cases; however, their KNN-based search cannot produce valid results for the remaining distributions. Neither HXT, gDel3D nor the Radiant Foam implementation supports the weighted case. We present results from the synthetic data in Fig.~\ref{fig:weighted_synthetic_cases_collage} and real-world data in Tab.~\ref{tab:timing_results_mipnerf360_weighted}. The results demonstrate that our method generalizes well to the weighted case across all datasets, achieving similar runtimes and scaling behavior as for the unweighted case. We provide further results on how our method scales with varying weight distributions in the supplementary material.

\subsection{Scaling mesh-based neural rendering}\label{sec:experiments:scaling_neural_rendering}
To demonstrate the practical benefits of efficient diagram construction at scale, we study the scaling behavior of a recent mesh-based neural rendering model, Radiant Foam~\cite{Govindarajan_2025_ICCV}. Radiant Foam represents scenes using differentiable 3D Voronoi diagrams and therefore requires frequent updates of cell adjacencies via Delaunay triangulation as the sites move during optimization. While the authors show that reconstruction quality improves consistently with an increasing number of points, indicating that higher geometric resolution leads to better reconstructions, their evaluation is limited to $2$M points.

In this context, the Voronoi diagram serves as a scene's spatial discretization. Each cell stores learnt parameters for density and appearance effects. Volumetric rendering is performed by casting rays through the diagram and accumulating the per-cell contributions, producing a final color for the pixel. During training, site positions and each site's attributes are optimized via gradient descent, which demands the mesh to be recomputed exhaustively, to reflect the changing geometry. The repeated reconstruction of the mesh is the computational bottleneck of scaling our method addresses. While Radiant Foam uses incremental updates when possible, we benchmark full reconstruction for a controlled comparison.

To enable Radiant Foam to scale beyond this regime, we replace their Delaunay triangulation step with our algorithm and retrain the model on scenes from the Mip-NeRF 360 dataset. We vary the final number of Voronoi sites from 3M to 17M, making no other changes to the method, training procedure, learning rate, or hyperparameters.

Fig.~\ref{fig:cover} reports the training time of Radiant Foam as a function of the number of points, along with LPIPS~\cite{zhang2018lpips} scores measured on the validation set as a function of training time for different final resolutions. The results highlight two key observations. First, mesh-based neural rendering models exhibit familiar scaling behavior, with reconstruction quality improving as the number of geometric primitives increases. Second, our approach enables such scaling with near-linear complexity, making large-scale Voronoi-based scene representations practically feasible.

\begin{table}[t]
    \small
    \centering
    \setlength{\tabcolsep}{1.0pt}
    \caption{\textbf{Power diagram creation/regular Delaunay triangulation}. Runtime results (in seconds) on point sets obtained from Radiant Foam checkpoints trained on the Mip-NeRF 360 dataset. Color coding denotes \colorbox{tbf}{fastest}, \colorbox{tbs}{second fastest}, and \colorbox{tbt}{third fastest}.}
    \Description{Power diagram / regular Delaunay runtimes in seconds across nine real-world scenes (1.9M--4.2M points) for CPU and GPU baselines, with the three fastest per scene color-coded. Our method is fastest on every scene on both GPUs while Basselin* fails on all of them; Geogram is the fastest CPU baseline.}
    \setlength{\tabcolsep}{2.5pt}
    \resizebox{1.0\linewidth}{!}{
    \begin{tabular}{cl ccccccccc}
    \toprule
    & & bicycle & bonsai & counter & drjohnson & garden & kitchen & playroom & room & stump \\
    & & 4.2M & 2.0M & 2.0M & 2.8M & 4.1M & 2.0M & 3.2M & 1.9M & 4.2M \\
    \midrule

    \multirow{4}{*}{\rotatebox{90}{CPU}} 
    & SciPy           & \cellcolor{tbn}212.538 & \cellcolor{tbn}94.919 & \cellcolor{tbn}117.140 & \cellcolor{tbn}174.597 & \cellcolor{tbn}230.981 & \cellcolor{tbn}124.567 & \cellcolor{tbn}188.740 & \cellcolor{tbn}110.533 & \cellcolor{tbn}214.884\\
    & CGAL            & \cellcolor{tbt}16.000 & \cellcolor{tbt}7.754 & \cellcolor{tbt}8.606 & \cellcolor{tbt}12.016 & \cellcolor{tbt}17.151 & \cellcolor{tbt}8.931 & \cellcolor{tbt}13.742 & \cellcolor{tbt}8.033 & \cellcolor{tbt}16.669\\
    & CGAL Parallel    & \cellcolor{tbn}- & \cellcolor{tbn}110.926 & \cellcolor{tbn}141.335 & \cellcolor{tbn}49.209 & \cellcolor{tbn}269.222 & \cellcolor{tbn}118.085 & \cellcolor{tbn}174.701 & \cellcolor{tbn}63.181 & \cellcolor{tbn}154.698\\
    & Geogram     & \cellcolor{tbs}2.939 & \cellcolor{tbs}1.368 & \cellcolor{tbs}1.444 & \cellcolor{tbs}1.776 & \cellcolor{tbs}2.670 & \cellcolor{tbs}1.379 & \cellcolor{tbs}2.666 & \cellcolor{tbs}1.321 & \cellcolor{tbs}3.206\\
    \midrule
    \multirow{2}{*}{\rotatebox{90}{5090}} 
    & Basselin*         & \cellcolor{tbn}- & \cellcolor{tbn}- & \cellcolor{tbn}- & \cellcolor{tbn}- & \cellcolor{tbn}- & \cellcolor{tbn}- & \cellcolor{tbn}- & \cellcolor{tbn}- & \cellcolor{tbn}-\\
    & Ours      & \cellcolor{tbf}1.134 & \cellcolor{tbf}0.319 & \cellcolor{tbf}0.266 & \cellcolor{tbf}0.346 & \cellcolor{tbf}0.613 & \cellcolor{tbf}0.248 & \cellcolor{tbf}0.385 & \cellcolor{tbf}0.264 & \cellcolor{tbf}0.836\\
    \midrule
    \multirow{2}{*}{\rotatebox{90}{H200}} 
    & Basselin*         & \cellcolor{tbn}- & \cellcolor{tbn}- & \cellcolor{tbn}- & \cellcolor{tbn}- & \cellcolor{tbn}- & \cellcolor{tbn}- & \cellcolor{tbn}- & \cellcolor{tbn}- & \cellcolor{tbn}-\\
    & Ours           & \cellcolor{tbf}1.697 & \cellcolor{tbf}0.827 & \cellcolor{tbf}0.414 & \cellcolor{tbf}0.419 & \cellcolor{tbf}1.038 & \cellcolor{tbf}0.465 & \cellcolor{tbf}0.462 & \cellcolor{tbf}0.344 & \cellcolor{tbf}1.280\\
    \bottomrule
    \end{tabular}
    }
    \label{tab:timing_results_mipnerf360_weighted}
\end{table}

\section{Conclusion}
We presented a GPU-based algorithm for efficiently and robustly constructing 3D Voronoi and power diagrams that supports large-scale point sets with complex spatial distributions while remaining competitive with state-of-the-art methods on small to moderate problem sizes.

Our approach is based on convex cell clipping, which naturally maps to massively parallel GPU execution. Combined with a culling criterion based on directional geometric bounds and a hierarchical best-first search strategy, this design yields an efficient and general algorithm that scales robustly with the number of points and across diverse spatial distributions.

We evaluated our method on both synthetic and real-world datasets, demonstrating performance comparable to or exceeding prior work on Delaunay triangulation across a wide range of distributions and scales. In several scenarios, our method achieves the best reported performance and naturally extends to the weighted case of power diagram construction.

Finally, we demonstrated the practical impact of our approach in the context of large-scale mesh-based neural rendering, where efficient and scalable construction of 3D Voronoi diagrams directly translates into improved reconstruction quality. These results underscore the importance of scalable Voronoi and power diagram algorithms as foundational tools for emerging large-scale graphics and vision applications.

In summary, we introduced a scalable and general method for 3D Voronoi and power diagram construction that matches or exceeds  prior Delaunay-based approaches and enables efficient processing of substantially larger point sets.

\paragraph{Limitations and Future Work.}
Our current implementation targets single-GPU execution and static point sets, and its scalability is ultimately bounded by available device memory rather than computation. Supporting incremental updates to diagrams, multi-GPU execution, and distributed construction remains an important direction for future work. In addition, our method relies on floating-point arithmetic and does not target the exact geometric guarantees of CPU-based libraries. Finally, further study of extreme weight distributions could provide additional insight into the limits of our culling strategies.

\begin{acks}

This work was partially supported by the Wallenberg AI, Autonomous Systems and Software Program (WASP) funded by the Knut and Alice Wallenberg Foundation. Computational resources were provided by NAISS at \href{https://www.nsc.liu.se/}{NSC Berzelius}, partially funded by the Swedish Research Council, grant agreement no. 2022-06725.

\end{acks}

\bibliographystyle{ACM-Reference-Format}
\bibliography{main}

@String{Computing = "Computing" }

@String{Computer = "{IEEE} Computer" }

@String{Springer = "Springer-Verlag" }

@article{Bower81,
    author = {Bowyer, A.},
    title = {Computing Dirichlet tessellations*},
    journal = {The Computer Journal},
    volume = {24},
    number = {2},
    pages = {162-166},
    year = {1981},
    month = {01},
    issn = {0010-4620},
    doi = {10.1093/comjnl/24.2.162},
    url = {https://doi.org/10.1093/comjnl/24.2.162},
    eprint = {https://academic.oup.com/comjnl/article-pdf/24/2/162/967239/240162.pdf},
}

@article{Watson81,
    author = {Watson, D. F.},
    title = {Computing the n-dimensional Delaunay tessellation with application to Voronoi polytopes*},
    journal = {The Computer Journal},
    volume = {24},
    number = {2},
    pages = {167-172},
    year = {1981},
    month = {01},
    issn = {0010-4620},
    doi = {10.1093/comjnl/24.2.167},
    url = {https://doi.org/10.1093/comjnl/24.2.167},
    eprint = {https://academic.oup.com/comjnl/article-pdf/24/2/167/967258/240167.pdf},
}

@article{aurenhammer1987power,
  title={Power diagrams: properties, algorithms and applications},
  author={Aurenhammer, Franz},
  journal={SIAM journal on computing},
  volume={16},
  number={1},
  pages={78--96},
  year={1987},
  publisher={SIAM}
}

@article{QHull96,
author = {Barber, C. Bradford and Dobkin, David P. and Huhdanpaa, Hannu},
title = {The quickhull algorithm for convex hulls},
year = {1996},
issue_date = {Dec. 1996},
publisher = {Association for Computing Machinery},
address = {New York, NY, USA},
volume = {22},
number = {4},
issn = {0098-3500},
url = {https://doi.org/10.1145/235815.235821},
doi = {10.1145/235815.235821},
journal = {ACM Trans. Math. Softw.},
month = dec,
pages = {469–483},
numpages = {15}
}

@article{rycroft2009voro++,
  title={VORO++: A three-dimensional Voronoi cell library in C++},
  author={Rycroft, Chris},
  year={2009}
}

@article{ParDel03,
author = {Chrisochoides, Nikos and Nave, Démian},
title = {Parallel Delaunay mesh generation kernel},
journal = {International Journal for Numerical Methods in Engineering},
volume = {58},
number = {2},
pages = {161-176},
doi = {https://doi.org/10.1002/nme.765},
url = {https://onlinelibrary.wiley.com/doi/abs/10.1002/nme.765},
eprint = {https://onlinelibrary.wiley.com/doi/pdf/10.1002/nme.765},
year = {2003}
}

@article{voro2023extension,
  title={An extension to Voro++ for multithreaded computation of Voronoi cells},
  author={Lu, Jiayin and Lazar, Emanuel A and Rycroftz, Chris H},
  journal={Computer Physics Communications},
  volume={291},
  pages={108832},
  year={2023},
  publisher={Elsevier}
}

@article{MeshlessVor18,
author = {Ray, Nicolas and Sokolov, Dmitry and Lefebvre, Sylvain and L\'{e}vy, Bruno},
title = {Meshless voronoi on the GPU},
year = {2018},
issue_date = {December 2018},
publisher = {Association for Computing Machinery},
address = {New York, NY, USA},
volume = {37},
number = {6},
issn = {0730-0301},
url = {https://doi.org/10.1145/3272127.3275092},
doi = {10.1145/3272127.3275092},
journal = {ACM Trans. Graph.},
month = dec,
articleno = {265},
numpages = {12}
}

@incollection{levy2013variational,
  title     = {Variational anisotropic surface meshing with Voronoi parallel linear enumeration},
  author    = {L{\'e}vy, Bruno and Bonneel, Nicolas},
  booktitle = {Proceedings of the 21st international meshing roundtable},
  pages     = {349--366},
  year      = {2013},
  publisher = {Springer}
}

@inproceedings{sainlot2017restricting,
  title        = {Restricting Voronoi diagrams to meshes using corner validation},
  author       = {Sainlot, Maxime and Nivoliers, Vincent and Attali, Dominique},
  booktitle    = {Computer Graphics Forum},
  volume       = {36},
  number       = {5},
  pages        = {81--91},
  year         = {2017},
  organization = {Wiley Online Library}
}

@article{marot2019one,
  title     = {One machine, one minute, three billion tetrahedra},
  author    = {Marot, C{\'e}lestin and Pellerin, Jeanne and Remacle, Jean-Fran{\c{c}}ois},
  journal   = {International Journal for Numerical Methods in Engineering},
  volume    = {117},
  number    = {9},
  pages     = {967--990},
  year      = {2019},
  publisher = {Wiley Online Library}
}

@InProceedings{Govindarajan_2025_ICCV,
    author    = {Govindarajan, Shrisudhan and Rebain, Daniel and Yi, Kwang Moo and Tagliasacchi, Andrea},
    title     = {Radiant Foam: Real-Time Differentiable Ray Tracing},
    booktitle = {Proceedings of the IEEE/CVF International Conference on Computer Vision (ICCV)},
    month     = {October},
    year      = {2025},
    pages     = {4135-4145}
}

@misc{mai2025radiancemeshesvolumetricreconstruction,
      title={Radiance Meshes for Volumetric Reconstruction}, 
      author={Alexander Mai and Trevor Hedstrom and George Kopanas and Janne Kontkanen and Falko Kuester and Jonathan T. Barron},
      year={2025},
      eprint={2512.04076},
      archivePrefix={arXiv},
      primaryClass={cs.GR},
      url={https://arxiv.org/abs/2512.04076}, 
}

@inproceedings{barron2022mip,
  title={Mip-nerf 360: Unbounded anti-aliased neural radiance fields},
  author={Barron, Jonathan T and Mildenhall, Ben and Verbin, Dor and Srinivasan, Pratul P and Hedman, Peter},
  booktitle={Proceedings of the IEEE/CVF conference on computer vision and pattern recognition},
  pages={5470--5479},
  year={2022}
}

@inproceedings{cao2014gpu,
  title={A GPU accelerated algorithm for 3D Delaunay triangulation},
  author={Cao, Thanh-Tung and Nanjappa, Ashwin and Gao, Mingcen and Tan, Tiow-Seng},
  booktitle={Proceedings of the 18th meeting of the ACM SIGGRAPH Symposium on Interactive 3D Graphics and Games},
  pages={47--54},
  year={2014}
}

@article{10.1111:cgf.142610,
    journal = {Computer Graphics Forum},
    title = {{Restricted Power Diagrams on the GPU}},
    author = {Basselin, Justine and Alonso, Laurent and Ray, Nicolas and Sokolov, Dmitry and Lefebvre, Sylvain and Lévy, Bruno},
    year = {2021},
    publisher = {The Eurographics Association and John Wiley & Sons Ltd.},
    ISSN = {1467-8659},
    DOI = {10.1111/cgf.142610}
}

@ARTICLE{2020SciPy-NMeth,
  author  = {Virtanen, Pauli and Gommers, Ralf and Oliphant, Travis E. and
            Haberland, Matt and Reddy, Tyler and Cournapeau, David and
            Burovski, Evgeni and Peterson, Pearu and Weckesser, Warren and
            Bright, Jonathan and {van der Walt}, St{\'e}fan J. and
            Brett, Matthew and Wilson, Joshua and Millman, K. Jarrod and
            Mayorov, Nikolay and Nelson, Andrew R. J. and Jones, Eric and
            Kern, Robert and Larson, Eric and Carey, C J and
            Polat, {\.I}lhan and Feng, Yu and Moore, Eric W. and
            {VanderPlas}, Jake and Laxalde, Denis and Perktold, Josef and
            Cimrman, Robert and Henriksen, Ian and Quintero, E. A. and
            Harris, Charles R. and Archibald, Anne M. and
            Ribeiro, Ant{\^o}nio H. and Pedregosa, Fabian and
            {van Mulbregt}, Paul and {SciPy 1.0 Contributors}},
  title   = {{{SciPy} 1.0: Fundamental Algorithms for Scientific
            Computing in Python}},
  journal = {Nature Methods},
  year    = {2020},
  volume  = {17},
  pages   = {261--272},
  adsurl  = {https://rdcu.be/b08Wh},
  doi     = {10.1038/s41592-019-0686-2},
}

@book{cgal:eb-25b,
  title = {{CGAL} User and Reference Manual},
  author = {{The CGAL Project}},
  publisher = {{CGAL Editorial Board}},
  edition = {{6.1}},
  year = 2025,
  url = {https://doc.cgal.org/6.1/Manual/packages.html}
}

@InProceedings{zhang2018lpips,
author = {Zhang, Richard and Isola, Phillip and Efros, Alexei A. and Shechtman, Eli and Wang, Oliver},
title = {The Unreasonable Effectiveness of Deep Features as a Perceptual Metric},
booktitle = {Proceedings of the IEEE Conference on Computer Vision and Pattern Recognition (CVPR)},
month = {June},
year = {2018}
}

@article{mildenhall2021nerf,
  title={{NeRF}: Representing scenes as neural radiance fields for view synthesis},
  author={Mildenhall, Ben and Srinivasan, Pratul P and Tancik, Matthew and Barron, Jonathan T and Ramamoorthi, Ravi and Ng, Ren},
  journal={Communications of the ACM},
  volume={65},
  number={1},
  pages={99--106},
  year={2021},
  publisher={ACM New York, NY, USA}
}

@Article{kerbl20233Dgaussians,
      author       = {Kerbl, Bernhard and Kopanas, Georgios and Leimk{\"u}hler, Thomas and Drettakis, George},
      title        = {{3D} {Gaussian} Splatting for Real-Time Radiance Field Rendering},
      journal      = {ACM Transactions on Graphics},
      number       = {4},
      volume       = {42},
      month        = {July},
      year         = {2023},
      url          = {https://repo-sam.inria.fr/fungraph/3d-gaussian-splatting/}
}

@article{wang2021neus,
  title={Neus: Learning neural implicit surfaces by volume rendering for multi-view reconstruction},
  author={Wang, Peng and Liu, Lingjie and Liu, Yuan and Theobalt, Christian and Komura, Taku and Wang, Wenping},
  journal={arXiv preprint arXiv:2106.10689},
  year={2021}
}

@article{yariv2021volume,
  title={Volume rendering of neural implicit surfaces},
  author={Yariv, Lior and Gu, Jiatao and Kasten, Yoni and Lipman, Yaron},
  journal={Advances in neural information processing systems},
  volume={34},
  pages={4805--4815},
  year={2021}
}

@inproceedings{yariv2023bakedsdf,
  title={Bakedsdf: Meshing neural sdfs for real-time view synthesis},
  author={Yariv, Lior and Hedman, Peter and Reiser, Christian and Verbin, Dor and Srinivasan, Pratul P and Szeliski, Richard and Barron, Jonathan T and Mildenhall, Ben},
  booktitle={ACM SIGGRAPH 2023 conference proceedings},
  pages={1--9},
  year={2023}
}

@inproceedings{choi2024meshgs,
  title={Meshgs: Adaptive mesh-aligned gaussian splatting for high-quality rendering},
  author={Choi, Jaehoon and Lee, Yonghan and Lee, Hyungtae and Kwon, Heesung and Manocha, Dinesh},
  booktitle={Proceedings of the Asian Conference on Computer Vision},
  pages={3310--3326},
  year={2024}
}

@article{gao2024real,
  title={Real-time large-scale deformation of gaussian splatting},
  author={Gao, Lin and Yang, Jie and Zhang, Bo-tao and Sun, Jia-mu and Yuan, Yu-jie and Fu, Hongbo and Lai, Yu-kun},
  journal={ACM Transactions on Graphics (TOG)},
  volume={43},
  number={6},
  pages={1--17},
  year={2024},
  publisher={ACM New York, NY, USA}
}

@article{lin2024direct,
  title={Direct learning of mesh and appearance via 3d gaussian splatting},
  author={Lin, Ancheng and Xiang, Yusheng and Kennedy, Paul and Li, Jun},
  journal={arXiv preprint arXiv:2405.06945},
  year={2024}
}

@article{aurenhammer1991voronoi,
  title={Voronoi diagrams—a survey of a fundamental geometric data structure},
  author={Aurenhammer, Franz},
  journal={ACM computing surveys (CSUR)},
  volume={23},
  number={3},
  pages={345--405},
  year={1991},
  publisher={ACM New York, NY, USA}
}

@incollection{brochu2010matching,
  title={Matching fluid simulation elements to surface geometry and topology},
  author={Brochu, Tyson and Batty, Christopher and Bridson, Robert},
  booktitle={ACM SIGGRAPH 2010 papers},
  pages={1--9},
  year={2010}
}

@article{shewchuk2002delaunay,
  title={Delaunay refinement algorithms for triangular mesh generation},
  author={Shewchuk, Jonathan Richard},
  journal={Computational geometry},
  volume={22},
  number={1-3},
  pages={21--74},
  year={2002},
  publisher={Elsevier}
}

@misc{gdel3dgithub,
  author       = {Nanjappa, Ashwin },
  title        = {gDel3D},
  year         = {2026},
  publisher    = {GitHub},
  journal      = {GitHub repository},
  howpublished = {\url{https://github.com/ashwin/gDel3D}},
  commit       = {a429be8d702a1f0f0d9366e14f7f492b1c071c9b}
}

@misc{radfoamgithub,
  author       = {Govindarajan, Shrisudhan and Rebain, Daniel},
  title        = {radfoam},
  year         = {2026},
  publisher    = {GitHub},
  journal      = {GitHub repository},
  howpublished = {\url{https://github.com/theialab/radfoam}},
  commit       = {3e7b52cf74e37ab2ab5e695f53570f515f537e3d}
}

@inproceedings{schoenberger2016sfm,
    author={Sch\"{o}nberger, Johannes Lutz and Frahm, Jan-Michael},
    title={Structure-from-Motion Revisited},
    booktitle={Conference on Computer Vision and Pattern Recognition (CVPR)},
    year={2016},
}

@article{marot2020quality,
  title={Quality tetrahedral mesh generation with HXT},
  author={Marot, C{\'e}lestin and Remacle, Jean-Fran{\c{c}}ois},
  journal={arXiv preprint arXiv:2008.08508},
  year={2020}
}

@misc{geogramgithub,
  author = {Lévy, Bruno},
  title = {Geogram: A Programming Library with Geometric Algorithms},
  year = {2025},
  howpublished = {\url{https://github.com/BrunoLevy/geogram}},
  note = {Accessed: 2025}
}

@article{hedman2018deep,
  title={Deep blending for free-viewpoint image-based rendering},
  author={Hedman, Peter and Philip, Julien and Price, True and Frahm, Jan-Michael and Drettakis, George and Brostow, Gabriel},
  journal={ACM Transactions on Graphics (ToG)},
  volume={37},
  number={6},
  pages={1--15},
  year={2018},
  publisher={ACM New York, NY, USA}
}

\begin{figure*}[tb]
    \centering
    \includegraphics[width=1\linewidth]{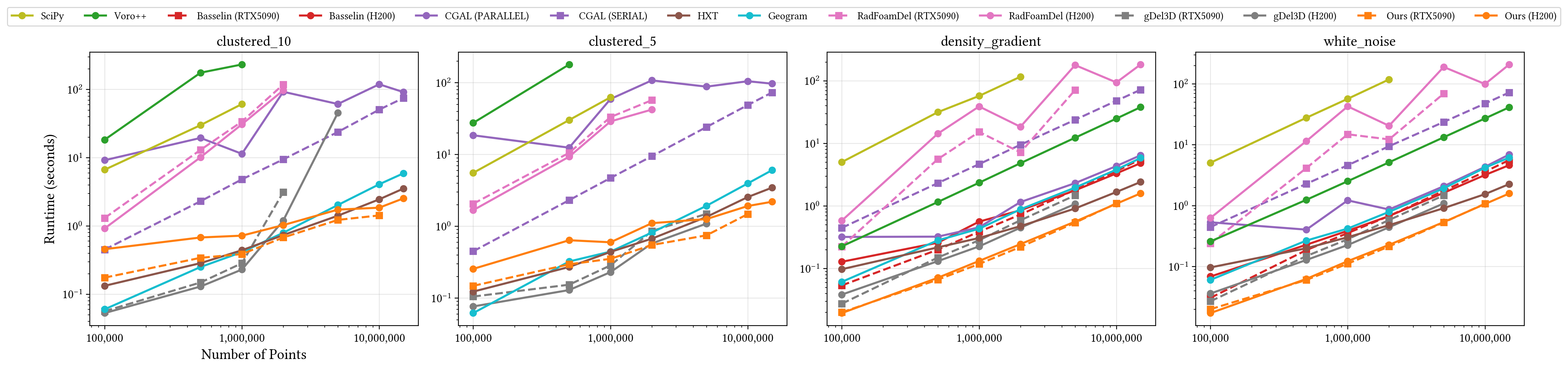}
    \caption{Runtime results (in seconds) for Voronoi diagram creation/Delaunay triangulation on the synthetic generated point sets described in Sec.~\ref{sec:experiments:datasets}. Missing data points indicate that the method could not finish within the maximum allowed time limit of 300 seconds.}
    \Description{Log-scale line plots of runtime versus point count (0.1M--15M) for CPU and GPU baselines across four synthetic distributions (white noise, density gradient, clustered K=5, clustered K=10) on RTX 5090 and H200. Our method's curve sits at or near the lowest runtime across all distributions, while several baselines fail or hit the 300-second timeout at large point counts.}
    \label{fig:synthetic_cases_collage}
\end{figure*}
\begin{figure*}[tb]
    \centering
    \includegraphics[width=1\linewidth]{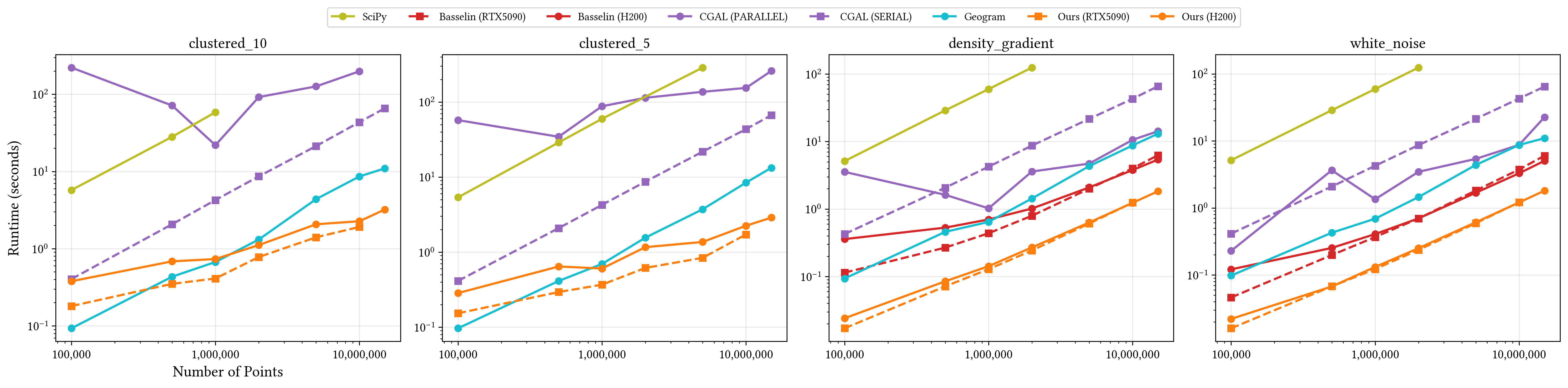}
    \caption{Runtime results (in seconds) for power diagram creation/regular Delaunay triangulation on the weighted synthetic generated point sets described in Sec.~\ref{sec:experiments:datasets}. Missing data points indicate that the method could not finish within the maximum allowed time limit of 300 seconds.}
    \Description{Log-scale line plots of runtime versus point count (0.1M--15M) for CPU and GPU power diagram methods across four weighted synthetic distributions on RTX 5090 and H200. Our method achieves the lowest runtime across all distributions, while CPU baselines grow steeply and the Basselin baseline only produces results on Poisson-distributed cases.}
    \label{fig:weighted_synthetic_cases_collage}
\end{figure*}
\begin{figure*}
    \centering
    \includegraphics[width=1\linewidth]{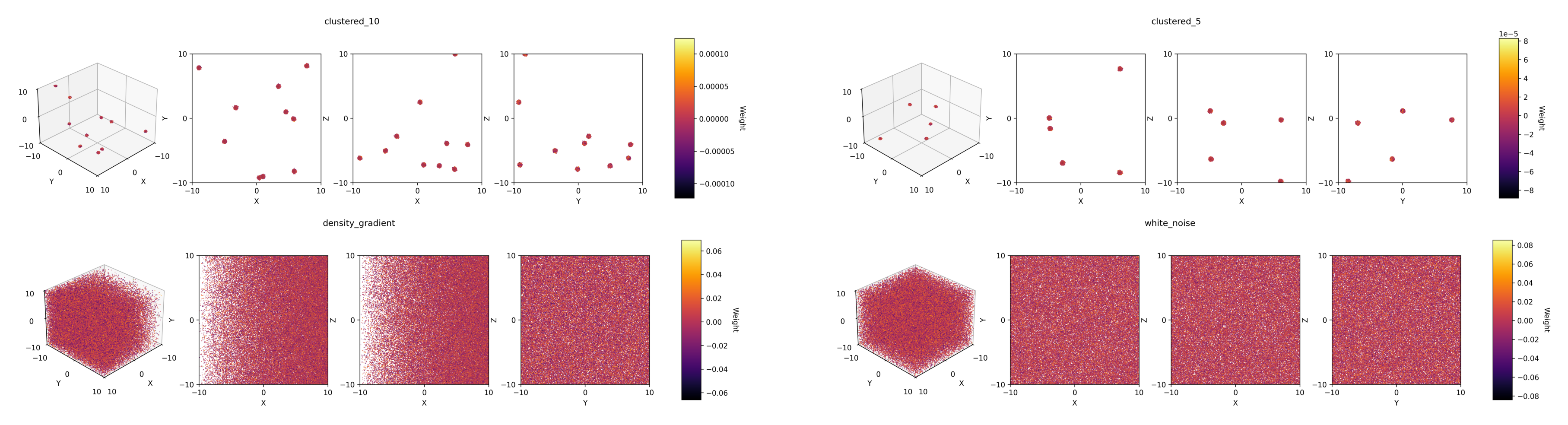}
    \caption{Visualization of synthetic test cases.}
    \Description{Four 3D scatter plots of synthetic point sets: a uniform white-noise distribution, a linear density gradient along one axis, and clustered distributions with five and ten Gaussian clusters. The visualizations contrast homogeneous and highly non-uniform spatial distributions.}
    \label{fig:visualization_synthetic_test_cases}
\end{figure*}

\clearpage 
\clearpage
\setcounter{page}{1}
\appendix
\section{Implementation details}
\begin{figure}
    \centering
    \includegraphics[width=0.9\linewidth]{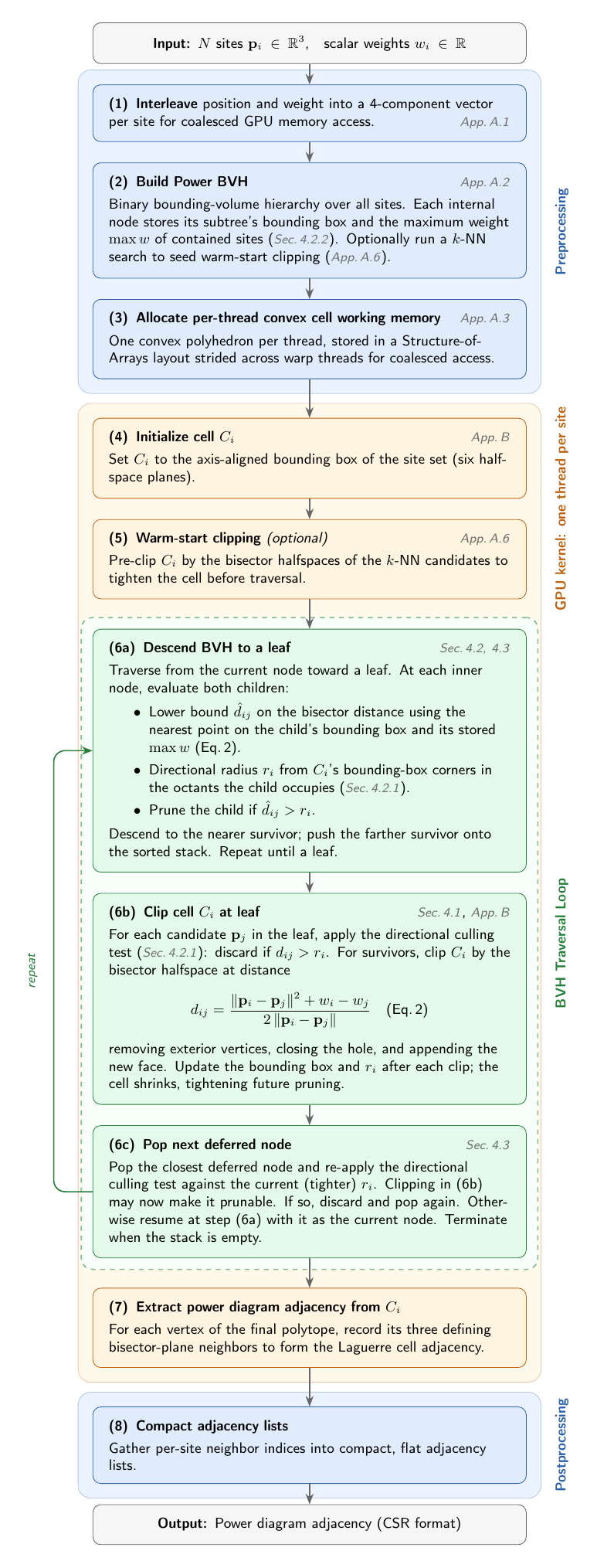}
    \caption{High-level diagram of our algorithm.}
    \Description{Flow diagram of our GPU pipeline: input weighted points are packed and indexed in a power-augmented BVH, then a fused kernel runs per-site best-first traversal, directional culling, and convex cell clipping to produce the final power diagram.}
    \label{fig:our_pipeline}
\end{figure}

Our method is implemented directly in CUDA, enabling the low-level control necessary for hardware-level performance optimization. Power diagram construction over $N$ sites decomposes naturally into $N$ independent volume-clipping tasks, which we map one-to-one onto GPU threads. Each thread executes a sequence of clipping operations that progressively refine the Voronoi cell geometry and record the set of neighboring sites. This parallel formulation scales directly with the number of threads, but achieving peak throughput demands careful attention to memory layout and data-access patterns, which we detail in the following subsections. We provide a high-level overview of our algorithm in Fig.~\ref{fig:our_pipeline} and detail its components below.

\subsection{Input packing}
Given $N$ input sites, each with position $\mathbf{p}_i \in \mathbb{R}^3$ and scalar weight $w_i \in \mathbb{R}$, we pack both in a continuous \texttt{float4}, aligning each site to a 128-bit boundary. This allows each site to be fetched with a single \texttt{LDG.128} instruction, and ensures that consecutive threads within a warp access consecutive \texttt{float4} elements, enabling memory coalescing and maximizing effective bandwidth.

\subsection{Power-augmented BVH}
We build a binary BVH over the site positions using the cuBQL GPU builder, configured with leaf size $\ell$ set to 17 for H200 and 10 for RTX5090. The BVH is augmented with a per-node $\texttt{maxweight}$ field, which stores the maximum weight in the subtree rooted at that node. This $\texttt{maxweight}$ is required to perform the culling criteria described in Sec.~\ref{sec:culling} during the hierarchical neighbor search.

\subsection{Per-thread memory}\label{sec:per_thread_memory}
Our clipping procedure builds on the work of \cite{10.1111:cgf.142610} but has been substantially adapted to serve our goal of computing the full connectivity of the power diagram, as opposed to only evaluating volume integrals over its cells. This change, combined with the need for robust performance on highly non-homogeneous point distributions, required a fundamentally different per-thread memory strategy.

Convex cell clipping is primarily memory-bound: each clipping operation fetches a neighbor position from global memory and iterates over vertices and bisecting planes multiple times to recompute the resulting convex cell. In~\cite{10.1111:cgf.142610}, intermediate per-thread data is held in shared memory. Instead, we allocate it in coalesced global memory. While shared memory offers lower latency, we found the resulting per-thread footprint to severely limit streaming multiprocessor (SM) occupancy, reducing overall hardware utilization. Coalesced global memory allows more warps to reside on each SM, yielding higher effective throughput in practice.

Each thread maintains a compact representation of its convex polyhedron alongside the neighbor data needed for subsequent clipping steps. Bisecting planes are stored as \texttt{float4} plane equations. Vertices of the polyhedron are encoded as \texttt{uchar3} triples, each holding the indices of the three planes whose intersection defines that vertex. We additionally cache the explicit vertex positions in global memory rather than recomputing them each time they are needed. Since every vertex must be tested against each candidate bisecting plane, caching eliminates repeated intersection computations and substantially reduces arithmetic overhead. Storing these positions is feasible due to the choice of using global memory, where the available capacity is orders of magnitude larger than in shared memory.

\subsection{Garbage Collection}
As the volume is progressively clipped, the fixed-size array of bisecting planes accumulates entries, some of which no longer contribute to the polyhedron. When the array exceeds $85\%$ occupancy, we compact it by discarding fully clipped planes, reclaiming capacity for subsequent iterations without increasing total allocation.

Our approach differs fundamentally from that of \cite{10.1111:cgf.142610}, where the shared-memory arrays are never compacted. Instead, when an array overflows, the kernel is relaunched with larger array sizes, reattempting construction from scratch. This overflow-and-retry strategy is a primary limitation on non-uniform point distributions, where cells in dense regions may require visiting and storing far more neighbors than shared memory can accommodate, regardless of retry budget.

\subsection{Best-first cell neighbor traversal}
As described in Sec.~\ref{sec:hierarchical}, the hierarchical neighbor search maintains a per-thread stack of BVH nodes tagged with their distance to the seed site. Rather than keeping the stack sorted, new entries are simply appended and the ordering deferred to the pop operation, which performs a linear scan to locate the nearest element. Removal is then handled by copying the last stack entry into the vacated slot, avoiding any shifting. This is logically equivalent to maintaining a sorted stack but trades a small amount of extra work at pop time for significantly fewer writes at push time.

\subsection{KNN warm start}
Prior to BVH traversal, we optionally run a K-nearest-neighbor query ($K=8$) on the same BVH to obtain candidate neighbors for each site. These candidates are used to pre-clip the power cell before traversal, bounding its initial search radius to approximately the power distance to the K-th nearest neighbor rather than infinity. This allows the directional pruning criterion (Sec.~\ref{sec:culling}) to cull nodes from the very root of the tree, substantially reducing the number of nodes and leaf primitives visited per cell. This optimization is beneficial on GPUs with large L2 caches (e.g., NVIDIA Blackwell consumer GPUs such as the RTX 5090), where the KNN pass is cheap relative to the traversal savings it enables. On datacenter GPUs such as the H200, HBM bandwidth already yields highly efficient traversal, and the overhead of the additional pass is not recovered. The warm start is therefore disabled on such hardware.

\section{Convex cell clipping}
In this section we describe the clipping procedure used in our algorithm.

\paragraph{Initialization}
As outlined in Sec.~\ref{sec:per_thread_memory}, each cell is represented by bisecting planes stored as half-space equations and vertex triplets holding the indices of the three half-spaces incident to that vertex.
For a given site $p_i$, the cell is initialized as the axis-aligned bounding box of the full point set. Candidate neighbors are then processed in order of proximity (as outlined in Secs.~\ref{sec:culling}–~\ref{sec:hierarchical}), progressively clipping the cell.

\paragraph{Clipping by a half-space}
A candidate neighbor $p_j$ induces a new half-space $h$ defined by the bisecting plane with $p_i$. The clipping proceeds in three steps. First, each vertex is classified by evaluating the new half-space equation. Vertices with a negative value lie outside the new half-space and are marked for removal. Second, marked vertices are removed one at a time, in an order that keeps each removed vertex adjacent to the current hole boundary. Removing vertices exposes a polygonal hole in the surface of the cell, whose boundary is maintained as a circular list of half-space indices. Each consecutive pair $(a,b)$ in the list represents a dangling edge lying along the intersection of face $a$ and face $b$. Each removal extends the boundary list by one edge, via a local update to the list. Third, the new facet is created, appending the new half-space $h$ to our half-space equations. For each consecutive pair $(a,b)$ in the circular hole boundary list, a new vertex triplet $(h, a, b)$ is added, closing the hole. If no vertices are marked for removal, the bisecting plane lies outside the current cell and no clipping is performed.

Finally, the bounding quantities of the cell are updated after clipping. 

\section{Baseline details}
We adopt the official implementation of gDel3D~\cite{gdel3dgithub} and report results using their default double-precision setup. We also evaluated gDel3D in single precision, but found the performance difference to be negligible (median < 0.5\%). For Radiant Foam~\cite{radfoamgithub}, we use the official implementation, which operates in single precision. Since the method of~\cite{10.1111:cgf.142610} was originally designed for volume integral computation rather than explicit mesh construction, we provide our own CUDA implementation of their algorithm to enable benchmarking, also in single precision. All CPU baselines are run in double precision.

\begin{table}%
    \small
    \centering
    \caption{Runtime (in seconds) for power diagram construction with varying weight distributions. Weights are sampled from a normal distribution with standard deviation proportional to the median nearest-neighbor distance, scaled by the weight ratio. The empty ratio denotes the fraction of cells with no neighboring cells.}
    \Description{Three-column table on the bicycle scene reporting weight ratio, empty-cell fraction, and runtime as the weight magnitude grows from 0 to 1e-1. Runtime stays near 0.6s for small ratios and rises to 2.24s as the empty-cell fraction reaches 0.47, showing graceful degradation with weight magnitude.}
    \begin{tabular}{lcc}
    \toprule
    Weight ratio & Empty ratio & Runtime (s) \\
    0.0 & 0.000 & 0.595 \\
    1e-6 & 0.000 & 0.596 \\
    1e-5 & 4.524e-6 & 0.596 \\
    1e-4 & 7.726e-4 & 0.598 \\
    1e-3 & 0.051 & 0.808 \\
    1e-2 & 0.250 & 1.469 \\
    1e-1 & 0.468 & 2.243 \\
    \bottomrule
    \end{tabular}
    \label{tab:power_weight_ablations}
\end{table}

\section{Further results}
In this section, we provide more detailed results of the runtime experiments described in Sec.~\ref{sec:experiments:results}. The results from the synthetic datasets are presented in Tab.~\ref{tab:timing_results_synthetic_clustered_10}, Tab.~\ref{tab:timing_results_synthetic_clustered_5},
Tab.~\ref{tab:timing_results_synthetic_density_gradient},
and Tab.~\ref{tab:timing_results_synthetic_white_noise} for the runs without weights, and in Tab.~\ref{tab:timing_results_synthetic_clustered_10_weighted}, Tab.~\ref{tab:timing_results_synthetic_clustered_5_weighted},
Tab.~\ref{tab:timing_results_synthetic_density_gradient_weighted},
and Tab.~\ref{tab:timing_results_synthetic_white_noise_weighted} for the runs with weights. Note that this is the same data as is visualized in Fig.~\ref{fig:synthetic_cases_collage} and Fig.~\ref{fig:weighted_synthetic_cases_collage}. Methods with missing results failed to complete the diagram construction, either by exceeding memory limits or by not finishing within the five-minute time limit.

\begin{table}[htbp]
    \small
    \centering
    \caption{Runtime results (in seconds) for Voronoi diagram creation/Delaunay triangulation on synthetic point sets created according to the clustered configuration described in Sec.~\ref{sec:experiments:datasets}, with $K=10$. Color coding denotes \colorbox{tbf}{fastest}, \colorbox{tbs}{second fastest}, and \colorbox{tbt}{third fastest}.}
    \Description{Voronoi/Delaunay runtimes in seconds on ten-cluster point sets from 0.1M to 15M points for CPU and GPU baselines, with the three fastest per size color-coded. gDel3D wins at small sizes but fails beyond 5M, where our method is fastest on the H200; Basselin* fails at every size.}
    \setlength{\tabcolsep}{2.5pt}
    \resizebox{\linewidth}{!}{
    \begin{tabular}{cl ccccccc}
    \toprule
    & & \multicolumn{7}{c}{10 clusters}\\
    & & \small 0.1M & \small 0.5M & \small 1.0M & \small 2.0M & \small 5.0M & \small 10.0M & \small 15.0M \\
    \midrule
    \multirow{5}{*}{\rotatebox{90}{CPU}}
    & SciPy                & \cellcolor{tbn}6.651 & \cellcolor{tbn}29.957 & \cellcolor{tbn}61.009 & \cellcolor{tbn}- & \cellcolor{tbn}- & \cellcolor{tbn}- & \cellcolor{tbn}-\\
    & Voro++         & \cellcolor{tbn}18.298 & \cellcolor{tbn}175.822 & \cellcolor{tbn}232.045 & \cellcolor{tbn}- & \cellcolor{tbn}- & \cellcolor{tbn}- & \cellcolor{tbn}-\\
    & CGAL                 & \cellcolor{tbt}0.449 & \cellcolor{tbt}2.299 & \cellcolor{tbt}4.808 & \cellcolor{tbt}9.431 & \cellcolor{tbt}23.789 & \cellcolor{tbt}50.547 & \cellcolor{tbt}74.723\\
    & CGAL Parallel      & \cellcolor{tbn}9.167 & \cellcolor{tbn}19.445 & \cellcolor{tbn}11.381 & \cellcolor{tbn}92.366 & \cellcolor{tbn}61.094 & \cellcolor{tbn}118.895 & \cellcolor{tbn}91.391\\
    & Geogram        & \cellcolor{tbf}0.060 & \cellcolor{tbf}0.250 & \cellcolor{tbf}0.413 & \cellcolor{tbs}0.792 & \cellcolor{tbs}2.024 & \cellcolor{tbs}4.072 & \cellcolor{tbs}5.890\\
    & HXT GmSH       & \cellcolor{tbs}0.132 & \cellcolor{tbs}0.285 & \cellcolor{tbs}0.440 & \cellcolor{tbf}0.733 & \cellcolor{tbf}1.410 & \cellcolor{tbf}2.450 & \cellcolor{tbf}3.520\\
    \midrule
    \multirow{4}{*}{\rotatebox{90}{5090}}
    & Basselin* & \cellcolor{tbn}- & \cellcolor{tbn}- & \cellcolor{tbn}- & \cellcolor{tbn}- & \cellcolor{tbn}- & \cellcolor{tbn}- & \cellcolor{tbn}-\\
    & RF Del        & \cellcolor{tbt}1.304 & \cellcolor{tbt}13.018 & \cellcolor{tbt}33.816 & \cellcolor{tbt}117.950 & \cellcolor{tbn}- & \cellcolor{tbn}- & \cellcolor{tbn}-\\
    & gDel3D        & \cellcolor{tbf}0.056 & \cellcolor{tbf}0.148 & \cellcolor{tbf}0.284 & \cellcolor{tbs}3.120 & \cellcolor{tbn}- & \cellcolor{tbn}- & \cellcolor{tbn}-\\
    & Ours    & \cellcolor{tbs}0.175 & \cellcolor{tbs}0.341 & \cellcolor{tbs}0.386 & \cellcolor{tbf}0.675 & \cellcolor{tbf}1.224 & \cellcolor{tbf}1.423 & \cellcolor{tbn}-\\
    \midrule
    \multirow{4}{*}{\rotatebox{90}{H200}}
    & Basselin* & \cellcolor{tbn}- & \cellcolor{tbn}- & \cellcolor{tbn}- & \cellcolor{tbn}- & \cellcolor{tbn}- & \cellcolor{tbn}- & \cellcolor{tbn}-\\
    & RF Del        & \cellcolor{tbt}0.917 & \cellcolor{tbt}10.136 & \cellcolor{tbt}30.805 & \cellcolor{tbt}96.942 & \cellcolor{tbn}- & \cellcolor{tbn}- & \cellcolor{tbn}-\\
    & gDel3D       & \cellcolor{tbf}0.053 & \cellcolor{tbf}0.129 & \cellcolor{tbf}0.229 & \cellcolor{tbs}1.196 & \cellcolor{tbs}45.423 & \cellcolor{tbn}- & \cellcolor{tbn}-\\
    & Ours    & \cellcolor{tbs}0.456 & \cellcolor{tbs}0.677 & \cellcolor{tbs}0.720 & \cellcolor{tbf}1.027 & \cellcolor{tbf}1.732 & \cellcolor{tbf}1.848 & \cellcolor{tbf}2.530\\
    \bottomrule
    \end{tabular}
    }
    \label{tab:timing_results_synthetic_clustered_10}
\end{table}

\begin{table}[htbp]
    \small
    \centering
    \caption{Runtime results (in seconds) for Voronoi diagram creation/Delaunay triangulation on synthetic point sets created according to the clustered configuration described in Sec.~\ref{sec:experiments:datasets}, with $K=5$. Color coding denotes \colorbox{tbf}{fastest}, \colorbox{tbs}{second fastest}, and \colorbox{tbt}{third fastest}.}
    \Description{Voronoi/Delaunay runtimes in seconds on five-cluster point sets from 0.1M to 15M points for CPU and GPU baselines, with the three fastest per size color-coded. gDel3D wins at smaller sizes but fails beyond 5M, where our method is fastest on the H200; Basselin* fails at every size.}
    \setlength{\tabcolsep}{2.5pt}
    \resizebox{\linewidth}{!}{
    \begin{tabular}{cl ccccccc}
    \toprule
    & & \multicolumn{7}{c}{5 clusters}\\
    & & \small 0.1M & \small 0.5M & \small 1.0M & \small 2.0M & \small 5.0M & \small 10.0M & \small 15.0M \\
    \midrule
    \multirow{5}{*}{\rotatebox{90}{CPU}}
    & SciPy                & \cellcolor{tbn}5.544 & \cellcolor{tbn}30.166 & \cellcolor{tbn}62.559 & \cellcolor{tbn}- & \cellcolor{tbn}- & \cellcolor{tbn}- & \cellcolor{tbn}-\\
    & Voro++         & \cellcolor{tbn}27.504 & \cellcolor{tbn}179.176 & \cellcolor{tbn}- & \cellcolor{tbn}- & \cellcolor{tbn}- & \cellcolor{tbn}- & \cellcolor{tbn}-\\
    & CGAL                 & \cellcolor{tbt}0.448 & \cellcolor{tbt}2.308 & \cellcolor{tbt}4.700 & \cellcolor{tbt}9.454 & \cellcolor{tbt}24.086 & \cellcolor{tbt}48.775 & \cellcolor{tbt}72.894\\
    & CGAL Parallel  & \cellcolor{tbn}18.486 & \cellcolor{tbn}12.434 & \cellcolor{tbn}59.192 & \cellcolor{tbn}107.589 & \cellcolor{tbn}87.969 & \cellcolor{tbn}104.733 & \cellcolor{tbn}96.761\\
    & Geogram        & \cellcolor{tbf}0.062 & \cellcolor{tbs}0.323 & \cellcolor{tbf}0.443 & \cellcolor{tbs}0.817 & \cellcolor{tbs}1.926 & \cellcolor{tbs}3.953 & \cellcolor{tbs}6.034\\
    & HXT GmSH       & \cellcolor{tbs}0.123 & \cellcolor{tbf}0.271 & \cellcolor{tbs}0.439 & \cellcolor{tbf}0.672 & \cellcolor{tbf}1.349 & \cellcolor{tbf}2.552 & \cellcolor{tbf}3.446\\
    \midrule
    \multirow{4}{*}{\rotatebox{90}{5090}}
    & Basselin* & \cellcolor{tbn}- & \cellcolor{tbn}- & \cellcolor{tbn}- & \cellcolor{tbn}- & \cellcolor{tbn}- & \cellcolor{tbn}- & \cellcolor{tbn}-\\
    & RF Del        & \cellcolor{tbt}2.055 & \cellcolor{tbt}10.536 & \cellcolor{tbt}33.169 & \cellcolor{tbt}56.991 & \cellcolor{tbn}- & \cellcolor{tbn}- & \cellcolor{tbn}-\\
    & gDel3D        & \cellcolor{tbf}0.105 & \cellcolor{tbf}0.153 & \cellcolor{tbf}0.285 & \cellcolor{tbs}0.857 & \cellcolor{tbs}1.489 & \cellcolor{tbn}- & \cellcolor{tbn}-\\
    & Ours    & \cellcolor{tbs}0.148 & \cellcolor{tbs}0.295 & \cellcolor{tbs}0.352 & \cellcolor{tbf}0.548 & \cellcolor{tbf}0.747 & \cellcolor{tbf}1.479 & \cellcolor{tbn}-\\
    \midrule
    \multirow{4}{*}{\rotatebox{90}{H200}}
    & Basselin* & \cellcolor{tbn}- & \cellcolor{tbn}- & \cellcolor{tbn}- & \cellcolor{tbn}- & \cellcolor{tbn}- & \cellcolor{tbn}- & \cellcolor{tbn}-\\
    & RF Del        & \cellcolor{tbt}1.674 & \cellcolor{tbt}9.309 & \cellcolor{tbt}28.851 & \cellcolor{tbt}42.143 & \cellcolor{tbn}- & \cellcolor{tbn}- & \cellcolor{tbn}-\\
    & gDel3D        & \cellcolor{tbf}0.076 & \cellcolor{tbf}0.129 & \cellcolor{tbf}0.230 & \cellcolor{tbf}0.576 & \cellcolor{tbf}1.088 & \cellcolor{tbn}- & \cellcolor{tbn}-\\
    & Ours    & \cellcolor{tbs}0.255 & \cellcolor{tbs}0.639 & \cellcolor{tbs}0.598 & \cellcolor{tbs}1.104 & \cellcolor{tbs}1.258 & \cellcolor{tbf}1.914 & \cellcolor{tbf}2.194\\
    \bottomrule
    \end{tabular}
    }
    \label{tab:timing_results_synthetic_clustered_5}
\end{table}

\begin{table}[htbp]
    \small
    \centering
    \caption{Runtime results (in seconds) for Voronoi diagram creation/Delaunay triangulation on synthetic point sets created according to the density gradient configuration described in Sec.~\ref{sec:experiments:datasets}. Color coding denotes \colorbox{tbf}{fastest}, \colorbox{tbs}{second fastest}, and \colorbox{tbt}{third fastest}.}
    \Description{Voronoi/Delaunay runtimes in seconds on density-gradient point sets from 0.1M to 15M points for CPU and GPU baselines, with the three fastest per size color-coded. Our method is the fastest GPU method at all completed sizes; gDel3D and SciPy fail at the largest sizes.}
    \setlength{\tabcolsep}{2.5pt}
    \resizebox{\linewidth}{!}{
    \begin{tabular}{cl ccccccc}
    \toprule
    & & \multicolumn{7}{c}{Density gradient}\\
    & & \small 0.1M & \small 0.5M & \small 1.0M & \small 2.0M & \small 5.0M & \small 10.0M & \small 15.0M \\
    \midrule
    \multirow{5}{*}{\rotatebox{90}{CPU}}
    & SciPy                & \cellcolor{tbn}5.019 & \cellcolor{tbn}31.462 & \cellcolor{tbn}57.210 & \cellcolor{tbn}116.067 & \cellcolor{tbn}- & \cellcolor{tbn}- & \cellcolor{tbn}-\\
    & Voro++         & \cellcolor{tbt}0.226 & \cellcolor{tbn}1.155 & \cellcolor{tbn}2.349 & \cellcolor{tbn}4.817 & \cellcolor{tbn}12.298 & \cellcolor{tbn}24.897 & \cellcolor{tbn}37.839\\
    & CGAL                 & \cellcolor{tbn}0.446 & \cellcolor{tbn}2.317 & \cellcolor{tbn}4.655 & \cellcolor{tbn}9.490 & \cellcolor{tbn}23.768 & \cellcolor{tbn}47.898 & \cellcolor{tbn}71.859\\
    & CGAL Parallel  & \cellcolor{tbn}0.320 & \cellcolor{tbt}0.323 & \cellcolor{tbt}0.455 & \cellcolor{tbt}1.154 & \cellcolor{tbt}2.313 & \cellcolor{tbt}4.337 & \cellcolor{tbt}6.455\\
    & Geogram        & \cellcolor{tbf}0.061 & \cellcolor{tbs}0.282 & \cellcolor{tbs}0.432 & \cellcolor{tbs}0.882 & \cellcolor{tbs}1.963 & \cellcolor{tbs}3.804 & \cellcolor{tbs}5.837\\
    & HXT GmSH       & \cellcolor{tbs}0.097 & \cellcolor{tbf}0.218 & \cellcolor{tbf}0.307 & \cellcolor{tbf}0.475 & \cellcolor{tbf}0.910 & \cellcolor{tbf}1.670 & \cellcolor{tbf}2.414\\
    \midrule
    \multirow{4}{*}{\rotatebox{90}{5090}}
    & Basselin*       & \cellcolor{tbt}0.054 & \cellcolor{tbt}0.200 & \cellcolor{tbt}0.387 & \cellcolor{tbt}0.728 & \cellcolor{tbt}1.810 & \cellcolor{tbs}3.602 & \cellcolor{tbf}5.574\\
    & RF Del               & \cellcolor{tbn}0.224 & \cellcolor{tbn}5.591 & \cellcolor{tbn}15.208 & \cellcolor{tbn}7.158 & \cellcolor{tbn}71.013 & \cellcolor{tbn}- & \cellcolor{tbn}-\\
    & gDel3D               & \cellcolor{tbs}0.027 & \cellcolor{tbs}0.149 & \cellcolor{tbs}0.283 & \cellcolor{tbs}0.590 & \cellcolor{tbs}1.476 & \cellcolor{tbn}- & \cellcolor{tbn}-\\
    & Ours           & \cellcolor{tbf}0.020 & \cellcolor{tbf}0.066 & \cellcolor{tbf}0.118 & \cellcolor{tbf}0.221 & \cellcolor{tbf}0.540 & \cellcolor{tbf}1.102 & \cellcolor{tbn}-\\
    \midrule
    \multirow{4}{*}{\rotatebox{90}{H200}}
    & Basselin*       & \cellcolor{tbt}0.128 & \cellcolor{tbt}0.258 & \cellcolor{tbt}0.561 & \cellcolor{tbt}0.852 & \cellcolor{tbt}1.773 & \cellcolor{tbs}3.311 & \cellcolor{tbs}4.829\\
    & RF Del               & \cellcolor{tbn}0.583 & \cellcolor{tbn}14.270 & \cellcolor{tbn}38.848 & \cellcolor{tbn}18.394 & \cellcolor{tbn}177.860 & \cellcolor{tbt}94.022 & \cellcolor{tbt}182.097\\
    & gDel3D               & \cellcolor{tbs}0.038 & \cellcolor{tbs}0.131 & \cellcolor{tbs}0.226 & \cellcolor{tbs}0.449 & \cellcolor{tbs}1.075 & \cellcolor{tbn}- & \cellcolor{tbn}-\\
    & Ours           & \cellcolor{tbf}0.019 & \cellcolor{tbf}0.071 & \cellcolor{tbf}0.132 & \cellcolor{tbf}0.245 & \cellcolor{tbf}0.561 & \cellcolor{tbf}1.084 & \cellcolor{tbf}1.588\\
    \bottomrule
    \end{tabular}
    }
    \label{tab:timing_results_synthetic_density_gradient}
\end{table}

\begin{table}[htbp]
    \small
    \centering
    \caption{Runtime results (in seconds) for Voronoi diagram creation/Delaunay triangulation on synthetic point sets created according to the uniform configuration described in Sec.~\ref{sec:experiments:datasets}. Color coding denotes \colorbox{tbf}{fastest}, \colorbox{tbs}{second fastest}, and \colorbox{tbt}{third fastest}.}
    \Description{Voronoi/Delaunay runtimes in seconds on uniformly sampled point sets from 0.1M to 15M points for CPU and GPU baselines, with the three fastest per size color-coded. Our method is fastest at all completed sizes; gDel3D and SciPy fail at the largest sizes.}
    \setlength{\tabcolsep}{2.5pt}
    \resizebox{\linewidth}{!}{
    \begin{tabular}{cl ccccccc}
    \toprule
    & & \multicolumn{7}{c}{White noise}\\
    & & \small 0.1M & \small 0.5M & \small 1.0M & \small 2.0M & \small 5.0M & \small 10.0M & \small 15.0M \\
    \midrule
    \multirow{5}{*}{\rotatebox{90}{CPU}}
    & SciPy                & \cellcolor{tbn}5.056 & \cellcolor{tbn}27.712 & \cellcolor{tbn}56.725 & \cellcolor{tbn}117.946 & \cellcolor{tbn}- & \cellcolor{tbn}- & \cellcolor{tbn}-\\
    & Voro++         & \cellcolor{tbt}0.258 & \cellcolor{tbn}1.246 & \cellcolor{tbn}2.518 & \cellcolor{tbn}5.138 & \cellcolor{tbn}13.239 & \cellcolor{tbn}27.031 & \cellcolor{tbn}41.136\\
    & CGAL                 & \cellcolor{tbn}0.444 & \cellcolor{tbn}2.280 & \cellcolor{tbn}4.603 & \cellcolor{tbn}9.436 & \cellcolor{tbn}23.611 & \cellcolor{tbn}47.415 & \cellcolor{tbn}71.999\\
    & CGAL Parallel  & \cellcolor{tbn}0.533 & \cellcolor{tbt}0.404 & \cellcolor{tbt}1.220 & \cellcolor{tbt}0.864 & \cellcolor{tbt}2.081 & \cellcolor{tbt}4.339 & \cellcolor{tbt}6.903\\
    & Geogram        & \cellcolor{tbf}0.060 & \cellcolor{tbs}0.267 & \cellcolor{tbs}0.420 & \cellcolor{tbs}0.791 & \cellcolor{tbs}1.920 & \cellcolor{tbs}4.202 & \cellcolor{tbs}6.138\\
    & HXT GmSH       & \cellcolor{tbs}0.097 & \cellcolor{tbf}0.200 & \cellcolor{tbf}0.303 & \cellcolor{tbf}0.478 & \cellcolor{tbf}0.909 & \cellcolor{tbf}1.545 & \cellcolor{tbf}2.259\\
    \midrule
    \multirow{4}{*}{\rotatebox{90}{5090}}
    & Basselin*       & \cellcolor{tbt}0.032 & \cellcolor{tbt}0.186 & \cellcolor{tbt}0.355 & \cellcolor{tbt}0.676 & \cellcolor{tbt}1.755 & \cellcolor{tbs}3.608 & \cellcolor{tbf}5.598\\
    & RF Del               & \cellcolor{tbn}0.240 & \cellcolor{tbn}4.139 & \cellcolor{tbn}14.837 & \cellcolor{tbn}12.218 & \cellcolor{tbn}68.997 & \cellcolor{tbn}- & \cellcolor{tbn}-\\
    & gDel3D               & \cellcolor{tbs}0.027 & \cellcolor{tbs}0.149 & \cellcolor{tbs}0.281 & \cellcolor{tbs}0.581 & \cellcolor{tbs}1.471 & \cellcolor{tbn}- & \cellcolor{tbn}-\\
    & Ours           & \cellcolor{tbf}0.020 & \cellcolor{tbf}0.060 & \cellcolor{tbf}0.112 & \cellcolor{tbf}0.214 & \cellcolor{tbf}0.530 & \cellcolor{tbf}1.078 & \cellcolor{tbn}-\\
    \midrule
    \multirow{4}{*}{\rotatebox{90}{H200}}
    & Basselin*       & \cellcolor{tbt}0.069 & \cellcolor{tbt}0.228 & \cellcolor{tbt}0.386 & \cellcolor{tbt}0.674 & \cellcolor{tbt}1.624 & \cellcolor{tbs}3.186 & \cellcolor{tbs}4.619\\
    & RF Del               & \cellcolor{tbn}0.627 & \cellcolor{tbn}11.486 & \cellcolor{tbn}42.595 & \cellcolor{tbn}20.458 & \cellcolor{tbn}188.883 & \cellcolor{tbt}99.114 & \cellcolor{tbt}207.671\\
    & gDel3D               & \cellcolor{tbs}0.036 & \cellcolor{tbs}0.128 & \cellcolor{tbs}0.225 & \cellcolor{tbs}0.442 & \cellcolor{tbs}1.086 & \cellcolor{tbn}- & \cellcolor{tbn}-\\
    & Ours           & \cellcolor{tbf}0.017 & \cellcolor{tbf}0.063 & \cellcolor{tbf}0.123 & \cellcolor{tbf}0.227 & \cellcolor{tbf}0.537 & \cellcolor{tbf}1.064 & \cellcolor{tbf}1.582\\
    \bottomrule
    \end{tabular}
    }
    \label{tab:timing_results_synthetic_white_noise}
\end{table}

\begin{table}[htbp]
    \small
    \centering
    \caption{Runtime results (in seconds) for power diagram creation/regular Delaunay triangulation on weighted synthetic point sets created according to the clustered configuration described in Sec.~\ref{sec:experiments:datasets}, with $K=10$. Color coding denotes \colorbox{tbf}{fastest}, \colorbox{tbs}{second fastest}, and \colorbox{tbt}{third fastest}.}
    \Description{Power diagram / regular Delaunay runtimes in seconds on weighted ten-cluster point sets from 0.1M to 15M points for CPU and GPU baselines, with the three fastest per size color-coded. Our method is fastest on GPU at every completed size while Basselin* fails on every size; Geogram is the fastest CPU baseline.}
    \setlength{\tabcolsep}{2.5pt}
    \resizebox{\linewidth}{!}{
    \begin{tabular}{cl ccccccc}
    \toprule
    & & \multicolumn{7}{c}{10 clusters}\\
    & & \small 0.1M & \small 0.5M & \small 1.0M & \small 2.0M & \small 5.0M & \small 10.0M & \small 15.0M \\
    \midrule
    \multirow{4}{*}{\rotatebox{90}{CPU}}
    & SciPy                & \cellcolor{tbt}5.737 & \cellcolor{tbt}28.115 & \cellcolor{tbt}58.569 & \cellcolor{tbn}- & \cellcolor{tbn}- & \cellcolor{tbn}- & \cellcolor{tbn}-\\
    & Voro++         & \cellcolor{tbt} & \cellcolor{tbn} & \cellcolor{tbn} & \cellcolor{tbn} & \cellcolor{tbn} & \cellcolor{tbn} & \cellcolor{tbn}\\
    & CGAL                 & \cellcolor{tbs}0.403 & \cellcolor{tbs}2.071 & \cellcolor{tbs}4.264 & \cellcolor{tbs}8.630 & \cellcolor{tbs}21.338 & \cellcolor{tbs}43.571 & \cellcolor{tbs}65.891\\
    & CGAL Parallel  & \cellcolor{tbn}221.661 & \cellcolor{tbn}71.534 & \cellcolor{tbn}21.930 & \cellcolor{tbt}92.072 & \cellcolor{tbt}126.464 & \cellcolor{tbt}198.626 & \cellcolor{tbn}-\\
    & Geogram        & \cellcolor{tbf}0.093 & \cellcolor{tbf}0.433 & \cellcolor{tbf}0.672 & \cellcolor{tbf}1.313 & \cellcolor{tbf}4.381 & \cellcolor{tbf}8.620 & \cellcolor{tbf}10.965\\
    \midrule
    \multirow{2}{*}{\rotatebox{90}{5090}}
    & Basselin* & \cellcolor{tbn}- & \cellcolor{tbn}- & \cellcolor{tbn}- & \cellcolor{tbn}- & \cellcolor{tbn}- & \cellcolor{tbn}- & \cellcolor{tbn}-\\
    & Ours     & \cellcolor{tbf}0.180 & \cellcolor{tbf}0.350 & \cellcolor{tbf}0.411 & \cellcolor{tbf}0.779 & \cellcolor{tbf}1.402 & \cellcolor{tbf}1.913 & \cellcolor{tbn}-\\
    \midrule
    \multirow{2}{*}{\rotatebox{90}{H200}}
    & Basselin*      & \cellcolor{tbn}- & \cellcolor{tbn}- & \cellcolor{tbn}- & \cellcolor{tbn}- & \cellcolor{tbn}- & \cellcolor{tbn}- & \cellcolor{tbn}-\\
    & Ours          & \cellcolor{tbf}0.379 & \cellcolor{tbf}0.685 & \cellcolor{tbf}0.734 & \cellcolor{tbf}1.113 & \cellcolor{tbf}2.071 & \cellcolor{tbf}2.267 & \cellcolor{tbf}3.210\\
    \bottomrule
    \end{tabular}
    }
    \label{tab:timing_results_synthetic_clustered_10_weighted}
\end{table}

\begin{table}[htbp]
    \small
    \centering
    \caption{Runtime results (in seconds) for power diagram creation/regular Delaunay triangulation on weighted synthetic point sets created according to the clustered configuration described in Sec.~\ref{sec:experiments:datasets}, with $K=5$. Color coding denotes \colorbox{tbf}{fastest}, \colorbox{tbs}{second fastest}, and \colorbox{tbt}{third fastest}.}
    \Description{Power diagram / regular Delaunay runtimes in seconds on weighted five-cluster point sets from 0.1M to 15M points for CPU and GPU baselines, with the three fastest per size color-coded. Our method is fastest on GPU at every completed size while Basselin* fails on every size; Geogram is the fastest CPU baseline.}
    \setlength{\tabcolsep}{2.5pt}
    \resizebox{\linewidth}{!}{
    \begin{tabular}{cl ccccccc}
    \toprule
    & & \multicolumn{7}{c}{5 clusters}\\
    & & \small 0.1M & \small 0.5M & \small 1.0M & \small 2.0M & \small 5.0M & \small 10.0M & \small 15.0M \\
    \midrule
    \multirow{4}{*}{\rotatebox{90}{CPU}}
    & SciPy                & \cellcolor{tbt}5.355 & \cellcolor{tbt}28.861 & \cellcolor{tbt}59.658 & \cellcolor{tbn}- & \cellcolor{tbn}287.128 & \cellcolor{tbn}- & \cellcolor{tbn}-\\
    & CGAL                 & \cellcolor{tbs}0.413 & \cellcolor{tbs}2.084 & \cellcolor{tbs}4.251 & \cellcolor{tbs}8.665 & \cellcolor{tbs}21.774 & \cellcolor{tbs}43.451 & \cellcolor{tbs}66.666\\
    & CGAL Parallel  & \cellcolor{tbn}57.310 & \cellcolor{tbn}34.397 & \cellcolor{tbn}87.747 & \cellcolor{tbt}113.751 & \cellcolor{tbt}136.586 & \cellcolor{tbt}153.878 & \cellcolor{tbt}259.233\\
    & Geogram        & \cellcolor{tbf}0.096 & \cellcolor{tbf}0.412 & \cellcolor{tbf}0.694 & \cellcolor{tbf}1.555 & \cellcolor{tbf}3.718 & \cellcolor{tbf}8.413 & \cellcolor{tbf}13.280\\
    \midrule
    \multirow{2}{*}{\rotatebox{90}{5090}}
    & Basselin* & \cellcolor{tbn}- & \cellcolor{tbn}- & \cellcolor{tbn}- & \cellcolor{tbn}- & \cellcolor{tbn}- & \cellcolor{tbn}- & \cellcolor{tbn}-\\
    & Ours     & \cellcolor{tbf}0.153 & \cellcolor{tbf}0.293 & \cellcolor{tbf}0.368 & \cellcolor{tbf}0.615 & \cellcolor{tbf}0.838 & \cellcolor{tbf}1.718 & \cellcolor{tbn}-\\
    \midrule
    \multirow{2}{*}{\rotatebox{90}{H200}}
    & Basselin* & \cellcolor{tbn}- & \cellcolor{tbn}- & \cellcolor{tbn}- & \cellcolor{tbn}- & \cellcolor{tbn}- & \cellcolor{tbn}- & \cellcolor{tbn}-\\
    & Ours     & \cellcolor{tbf}0.284 & \cellcolor{tbf}0.643 & \cellcolor{tbf}0.604 & \cellcolor{tbf}1.162 & \cellcolor{tbf}1.364 & \cellcolor{tbf}2.243 & \cellcolor{tbf}2.879\\
    \bottomrule
    \end{tabular}
    }
    \label{tab:timing_results_synthetic_clustered_5_weighted}
\end{table}

\begin{table}[htbp]
    \small
    \centering
    \caption{Runtime results (in seconds) for power diagram creation/regular Delaunay triangulation on weighted synthetic point sets created according to the density gradient configuration described in Sec.~\ref{sec:experiments:datasets}. Color coding denotes \colorbox{tbf}{fastest}, \colorbox{tbs}{second fastest}, and \colorbox{tbt}{third fastest}.}
    \Description{Power diagram / regular Delaunay runtimes in seconds on weighted density-gradient point sets from 0.1M to 15M points for CPU and GPU baselines, with the three fastest per size color-coded. Our method is fastest at all completed sizes; Basselin* is consistently second on GPUs and Geogram is the fastest CPU baseline.}
    \setlength{\tabcolsep}{2.5pt}
    \resizebox{\linewidth}{!}{
    \begin{tabular}{cl ccccccc}
    \toprule
    & & \multicolumn{7}{c}{Density gradient}\\
    & & \small 0.1M & \small 0.5M & \small 1.0M & \small 2.0M & \small 5.0M & \small 10.0M & \small 15.0M \\
    \midrule
    \multirow{4}{*}{\rotatebox{90}{CPU}}
    & SciPy                & \cellcolor{tbn}5.123 & \cellcolor{tbn}28.899 & \cellcolor{tbn}59.681 & \cellcolor{tbn}124.247 & \cellcolor{tbn}- & \cellcolor{tbn}- & \cellcolor{tbn}-\\
    & CGAL                 & \cellcolor{tbs}0.426 & \cellcolor{tbt}2.068 & \cellcolor{tbt}4.239 & \cellcolor{tbt}8.685 & \cellcolor{tbt}21.639 & \cellcolor{tbt}42.902 & \cellcolor{tbt}65.660\\
    & CGAL Parallel  & \cellcolor{tbt}3.545 & \cellcolor{tbs}1.623 & \cellcolor{tbs}1.023 & \cellcolor{tbs}3.589 & \cellcolor{tbs}4.707 & \cellcolor{tbs}10.567 & \cellcolor{tbs}14.202\\
    & Geogram        & \cellcolor{tbf}0.094 & \cellcolor{tbf}0.459 & \cellcolor{tbf}0.644 & \cellcolor{tbf}1.431 & \cellcolor{tbf}4.320 & \cellcolor{tbf}8.725 & \cellcolor{tbf}13.039\\
    \midrule
    \multirow{2}{*}{\rotatebox{90}{5090}}
    & Basselin* & \cellcolor{tbs}0.114 & \cellcolor{tbs}0.268 & \cellcolor{tbs}0.437 & \cellcolor{tbs}0.793 & \cellcolor{tbs}2.001 & \cellcolor{tbs}3.998 & \cellcolor{tbf}6.258\\
    & Ours     & \cellcolor{tbf}0.017 & \cellcolor{tbf}0.071 & \cellcolor{tbf}0.128 & \cellcolor{tbf}0.243 & \cellcolor{tbf}0.609 & \cellcolor{tbf}1.242 & \cellcolor{tbn}-\\
    \midrule
    \multirow{2}{*}{\rotatebox{90}{H200}}
    & Basselin* & \cellcolor{tbs}0.358 & \cellcolor{tbs}0.530 & \cellcolor{tbs}0.699 & \cellcolor{tbs}1.016 & \cellcolor{tbs}2.087 & \cellcolor{tbs}3.772 & \cellcolor{tbs}5.395\\
    & Ours     & \cellcolor{tbf}0.024 & \cellcolor{tbf}0.085 & \cellcolor{tbf}0.142 & \cellcolor{tbf}0.270 & \cellcolor{tbf}0.632 & \cellcolor{tbf}1.233 & \cellcolor{tbf}1.830\\
    \bottomrule
    \end{tabular}
    }
    \label{tab:timing_results_synthetic_density_gradient_weighted}
\end{table}

\begin{table}[htbp]
    \small
    \centering
    \setlength{\tabcolsep}{0.5pt}
    \caption{Runtime results (in seconds) for power diagram creation/regular Delaunay triangulation on weighted synthetic point sets created according to the uniform configuration described in Sec.~\ref{sec:experiments:datasets}. Color coding denotes \colorbox{tbf}{fastest}, \colorbox{tbs}{second fastest}, and \colorbox{tbt}{third fastest}.}
    \Description{Power diagram / regular Delaunay runtimes in seconds on weighted uniformly distributed point sets from 0.1M to 15M points for CPU and GPU baselines, with the three fastest per size color-coded. Our method is fastest at all completed sizes; Basselin* is consistently second on GPUs and Geogram is the fastest CPU baseline.}
    \setlength{\tabcolsep}{2.5pt}
    \resizebox{\linewidth}{!}{
    \begin{tabular}{cl ccccccc}
    \toprule
    & & \multicolumn{7}{c}{White noise}\\
    & & \small 0.1M & \small 0.5M & \small 1.0M & \small 2.0M & \small 5.0M & \small 10.0M & \small 15.0M \\
    \midrule
    \multirow{4}{*}{\rotatebox{90}{CPU}}
    & SciPy                & \cellcolor{tbn}5.151 & \cellcolor{tbn}28.726 & \cellcolor{tbn}59.514 & \cellcolor{tbn}124.243 & \cellcolor{tbn}- & \cellcolor{tbn}- & \cellcolor{tbn}-\\
    & CGAL                 & \cellcolor{tbt}0.413 & \cellcolor{tbs}2.087 & \cellcolor{tbt}4.262 & \cellcolor{tbt}8.657 & \cellcolor{tbt}21.354 & \cellcolor{tbt}42.925 & \cellcolor{tbt}64.724\\
    & CGAL Parallel  & \cellcolor{tbs}0.229 & \cellcolor{tbt}3.649 & \cellcolor{tbs}1.349 & \cellcolor{tbs}3.462 & \cellcolor{tbs}5.401 & \cellcolor{tbs}8.735 & \cellcolor{tbs}22.632\\
    & Geogram        & \cellcolor{tbf}0.098 & \cellcolor{tbf}0.427 & \cellcolor{tbf}0.692 & \cellcolor{tbf}1.454 & \cellcolor{tbf}4.384 & \cellcolor{tbf}8.712 & \cellcolor{tbf}11.003\\
    \midrule
    \multirow{2}{*}{\rotatebox{90}{5090}}
    & Basselin*  & \cellcolor{tbs}0.046 & \cellcolor{tbs}0.198 & \cellcolor{tbs}0.368 & \cellcolor{tbs}0.696 & \cellcolor{tbs}1.821 & \cellcolor{tbs}3.753 & \cellcolor{tbf}6.019\\
    & Ours      & \cellcolor{tbf}0.016 & \cellcolor{tbf}0.068 & \cellcolor{tbf}0.123 & \cellcolor{tbf}0.237 & \cellcolor{tbf}0.592 & \cellcolor{tbf}1.216 & \cellcolor{tbn}-\\
    \midrule
    \multirow{2}{*}{\rotatebox{90}{H200}}
    & Basselin*  & \cellcolor{tbs}0.121 & \cellcolor{tbs}0.253 & \cellcolor{tbs}0.408 & \cellcolor{tbs}0.698 & \cellcolor{tbs}1.686 & \cellcolor{tbs}3.294 & \cellcolor{tbs}5.053\\
    & Ours      & \cellcolor{tbf}0.022 & \cellcolor{tbf}0.067 & \cellcolor{tbf}0.132 & \cellcolor{tbf}0.251 & \cellcolor{tbf}0.612 & \cellcolor{tbf}1.208 & \cellcolor{tbf}1.811\\
    \bottomrule
    \end{tabular}
    }
    \label{tab:timing_results_synthetic_white_noise_weighted}
\end{table}

\paragraph{Timing breakdown}
The BVH construction in our method accounts for 4.8\% of the total time on large point sets. The remaining 95.2\% is our fused kernel, of which the vast majority of the time is spent on cell clipping operations (96.2\%) and the rest is BVH traversal and culling.

\paragraph{Precision}
We run our method using single precision. We compare the exactness of our method against the results from CGAL's exact predicates. On average, we note a small fraction of mismatching adjacencies (0.1–0.2\%) on real data, traced to near-coincident candidate planes. Volume rendering validation across 100k rays confirm negligible differences in accumulated density (mean < 1.7e-5, median $=0.0$), with consistent results on synthetic data.

\paragraph{Component ablations}
We ablate the two key algorithmic components of our method on a RTX 5090 and report results in Tab.~\ref{tab:component_ablations}. The first variant, \emph{isotropic culling}, replaces our directional culling criterion (Sec.~\ref{sec:culling}) with a single-radius bound, consistently degrading performance by $1.03$--$1.61\times$ across all datasets. The second variant, \emph{depth-first traversal}, replaces the best-first neighbor ordering (Sec.~\ref{sec:hierarchical}) with a standard depth-first BVH traversal. The impact is far more pronounced: without proximity-first ordering, the method clips against distant, non-contributing neighbors first, leading to slowdowns of up to $80\times$ on large clustered and real-world data. On Poisson-distributed data at small sizes the effect is mild, as the uniform density limits the penalty of suboptimal ordering.

\begin{table}%
    \small
    \centering
    \caption{Ablation study on the RTX 5090. We compare our full method against two variants: \emph{isotropic culling}, which replaces our directional criterion with a single-radius bound, and \emph{depth-first traversal}, which replaces best-first ordering with standard DFS. Reported values are the median slowdown factor (and range) relative to our full method.}
    \Description{Ablation table on the RTX 5090 reporting median slowdown factors (with ranges) of two variants relative to our full method across Clustered, Poisson, and Real-world datasets. Isotropic culling causes mild 1.26x--1.40x slowdowns, while depth-first traversal degrades up to 80x on non-uniform data and remains mild on Poisson.}
    \resizebox{\columnwidth}{!}{
    \begin{tabular}{lccc}
    \toprule
     & Clustered & Poisson & Real-world \\
    \midrule
    Isotropic culling      & 1.26$\times$ (1.03--1.52) & 1.33$\times$ (1.25--1.44) & 1.40$\times$ (1.34--1.61) \\
    Depth-first traversal  & 27.4$\times$ (3.96--80.3) & 1.30$\times$ (1.13--22.8) & 37.2$\times$ (34.8--53.1) \\
    \bottomrule
    \end{tabular}
    }
    \label{tab:component_ablations}
\end{table}

\paragraph{Varying weight distributions}
Additionally, we present runtime results for power diagram construction with varying weight distributions in Tab.~\ref{tab:power_weight_ablations}. We run our method on the bicycle scene from the real-world test cases described in Sec.~\ref{sec:experiments:datasets}, where cell weights are sampled from a normal distribution with standard deviation proportional to the median nearest-neighbor distance, scaled by a weight ratio. To provide intuition on the influence of the weight magnitude, we report the empty ratio, defined as the fraction of cells with no neighboring cells. The observed increase in runtime with larger weights is expected, as larger weights inflate the power distances between cells and ultimately increase the number of candidates that must be considered.

\end{document}